\documentclass[aps,showpacs,preprintnumbers,amsmath, amssymb]{revtex4}

\usepackage{float}
\usepackage{graphics,epsfig}
\usepackage{graphicx}
\usepackage{dcolumn}
\usepackage{bm}

\begin{document}
\baselineskip=0.8 cm

\begin{centering}
\vspace{1cm}
{\Large {\bf Holographic Superconductors with various condensates in Einstein-Gauss-Bonnet gravity
}}\\

\vspace{1.5cm}

 {\bf Qiyuan Pan $^{1,2}$}, {\bf Bin Wang $^{1}$}\\
 \vspace{.2in}
 $^{1}$ Department of Physics, Fudan University, 200433 Shanghai, China \\
$^{2}$ Institute of Physics and Department of Physics, Hunan Normal University, 410081 Changsha, China\\

\vspace{.2in}
 {\bf Eleftherios Papantonopoulos}\\
\vspace{.2in}
Department of Physics, National Technical University of Athens,\\
Zografou Campus GR 157 73, Athens, Greece
 \\

\vspace{.2in}
 {\bf Jeferson de  Oliveira}, {\bf A.B. Pavan}\\
\vspace{.2in} Instituto de Fisica, Universidade de Sao Paulo,\\ C.P.66.318, CEP 05315-970, Sao Paulo, Brazil
 \\

\vspace{3mm}

\end{centering}

\vspace*{0.2cm}
\begin{abstract}
\baselineskip=0.6 cm
\begin{center}
{\bf Abstract}
\end{center}

We study holographic superconductors in Einstein-Gauss-Bonnet
gravity. We consider two particular backgrounds: a $d$-dimensional
Gauss-Bonnet-AdS black hole and a Gauss-Bonnet-AdS soliton. We
discuss in detail the effects that the mass of the scalar field,
the Gauss-Bonnet coupling and the dimensionality of the AdS space
have on the condensation formation and conductivity. We also study
the ratio $\omega_g/T_c $ for various masses of the scalar field
and Gauss-Bonnet couplings.

\end{abstract}

\maketitle
\newpage
\vspace*{0.2cm}

\section{Introduction}

The AdS/CFT correspondence has become a powerful tool in studying strongly coupled phenomena in quantum field theory using results from a weakly
coupled gravity background. According to this correspondence principle~\cite{Maldacena,Witten}, a string theory on asymptotically AdS spacetimes
can be related to a conformal field theory on the boundary. In recent years, apart from string theory, this holographic correspondence, following
a more phenomenological approach, has also been applied to condensed matter physics and in particular to superconductivity.  It was first
suggested in \cite{Gubser:2005ih,GubserPRD78} that near the horizon of a charged black hole there is in operation a geometrical mechanism
parametrized by a charged scalar field of breaking a local $U(1)$ gauge symmetry. Then it was suggested to use the gauge/gravity duality to
construct gravitational duals of the transition from normal to superconducting states in the boundary theory \cite{GubserPRD78}.

The gravity dual of a superconductor consists of a system with a
black hole and a charged scalar field, in which the black hole
admits scalar hair at temperature smaller than a critical
temperature, while there is no scalar hair at larger
 temperatures~\cite{HartnollPRL101}. A condensate of the charged scalar field is
formed through its coupling to a Maxwell field of the background.
Neither field is backreacting on the metric.  Considering
fluctuations of the vector potential, the frequency dependent
conductivity was calculated, and it was shown that it develops a
gap determined by the condensate. This model  was further studied
beyond the probe limit~\cite{HartnollJHEP12}.   Along this line,
there have been a lot of investigations concerning the application
of AdS/CFT correspondence to condensed matter physics
\cite{HorowitzPRD78,Nakano-Wen,Amado,
Koutsoumbas,Maeda79,Sonner,Cai-Zhang}. See Refs.
\cite{HartnollRev,HerzogRev} for reviews.

These models however are phenomenological models. The classical
fields and their interactions are chosen by hand. It would have
also been desirable that these models emerge from a consistent
string theory~\cite{Ammon:2008fc,Gubser:2009qm,Gauntlett:2009dn}.
Also recently there is an effort to discuss stringy effects coming
from higher dimensions. The various condensates were studied in
(2+1) and in (3+1) superconductors \cite{HorowitzPRD78}. It was
found that there is a universal relation between the gap
$\omega_g$ in the frequency dependent conductivity and the
critical temperature $T_c$: $\omega_g/T_c \simeq 8$ respected to a
good approximation by all cases considered.

Motivated by the application of the Mermin-Wagner theorem to the holographic superconductors there was a study of the effects of the curvature
corrections on the $(3+1)$-dimensional superconductor \cite{Gregory}. A model of a charged scalar field together with a Maxwell field in the
five-dimensional Gauss-Bonnet-AdS black hole background was presented. It was found that higher curvature corrections make condensation harder
and cause the behavior of the ratio $\omega_g/T_c \simeq 8$ which was claimed to be universal in \cite{HorowitzPRD78} unstable. They presented a
semi-analytic approximation method to explain the qualitative features of superconductors giving fairly good agreement with numerical results.

In this work, we will carry out a detailed study in the probe limit of various condensates of holographic superconductors with curvature
corrections. We will consider two particular backgrounds: a $d$-dimensional Gauss-Bonnet-AdS black hole
\cite{Boulware-Deser,Cai-2002,Charmousis:2002rc} and a Gauss-Bonnet-AdS soliton \cite{Cai-Kim-Wang} background. Our aim is to clarify the
influence of the stringy effects on holographic superconductors in various dimensions.

In the case of $d$-dimensional Gauss-Bonnet-AdS black hole background, besides the study of the  influence the mass of the scalar field  and
Gauss-Bonnet coupling have on the formation of the scalar condensation, we will also present an analysis of the effects the dimensionality of the
AdS space have on the scalar condensation formation. We applied the semi-analytical method in matching the approximate solutions near the horizon
and the asymptotic AdS region \cite{Gregory} for $d\geq5$. We find that for $d>5$ the method breaks down unless the matching point is selected in
an appropriate range. We also study the ratio $\omega_g/T_c $ for various mass of the scalar field and Gauss-Bonnet coupling.

The AdS soliton is a gravitational configuration which has lower
energy than the AdS space in the Poincare coordinates, but has the
same boundary topology as the Ricci flat black hole and the AdS
space in the Poincare coordinates \cite{myers}.  It was found that
there is a Hawking-Page phase transition  between the Ricci flat
AdS black hole and the AdS soliton \cite{wit}. The signature of
this phase transition shows up in the quasi-normal modes spectrum
\cite{shen}.  More recently the Hawking-Page phase transition
between Ricci flat black holes and deformed AdS soliton in the
Gauss-Bonnet gravity was discussed in \cite{Cai-Kim-Wang}. It was
argued that although in Gauss-Bonnet gravity, the black hole
solution and AdS soliton are greatly deformed by the Gauss-Bonnet
term, the Gauss-Bonnet coefficient disappears in the Euclidean
action and as a result the Gauss-Bonnet term has no effect on the
Hawking-Page phase transition.



Recently it was found that there is a superconducting  phase dual
to an AdS soliton configuration \cite{Nishioka-Ryu-Takayanagi}. We
will extend the construction to include
  a Ricci flat AdS soliton
in Gauss-Bonnet gravity. We will study the effects of the mass of
the scalar field and the Gauss-Bonnet coupling on the scalar
condensation and conductivity and compare them with the
corresponding results in Gauss-Bonnet-AdS black hole
configuration.

The plan of the paper is the following. In Sec. 2 we present the basic equations of the holographic superconductor in the Gauss-Bonnet-AdS black
hole background. In Sec. 3 we explore the effects of the Gauss-Bonnet term, the spacetime dimension and the mass of the scalar field on the
scalar condensation and conductivity.  In Sec. 4 we discuss  the Gauss-Bonnet-AdS soliton background. We conclude in the last section with our
main results.

\section{Holographic Superconductors dual to Gauss-Bonnet-AdS black
holes}

The Einstein-Gauss-Bonnet theory is the most general Lovelock theory in five and six dimensions and the action with a negative cosmological
constant $\Lambda=-(d-1)(d-2)/2L^{2}$  is of the form
\begin{eqnarray}\label{BH action}
S=\frac{1}{16\pi G}\int
d^{d}x\sqrt{-g}\left[R+\frac{(d-1)(d-2)}{L^2}+\tilde{\alpha}
\left(R_{\mu\nu\gamma\delta}R^{\mu\nu\gamma\delta}-4R_{\mu\nu}R^{\mu\nu}+
R^{2}\right)\right] \;,
\end{eqnarray}
where $\tilde{\alpha}$ is the Gauss-Bonnet coupling constant with dimension $(length)^{2}$.  Considering that the Gauss-Bonnet term is an
effective string correction to gravity, the coupling $\tilde{\alpha}$ is connected to the string coupling, therefore it is positive. The
background solution of a neutral black hole is described by  \cite{Boulware-Deser,Cai-2002,Charmousis:2002rc}
\begin{eqnarray}\label{BH metric}
ds^2=-f(r)dt^{2}+\frac{dr^2}{f(r)}+r^{2}(d\chi^2+h_{ij}dx^{i}dx^{j})~,
\end{eqnarray}
where
\begin{eqnarray}
f(r)=\frac{r^2}{2\alpha}\left[1-\sqrt{1+\frac{64\pi G\alpha
M}{(d-2) \Sigma r^{d-1}}-\frac{4\alpha}{L^{2}}}~\right]\label{fr}.
\end{eqnarray}
Here $\alpha=\tilde{\alpha}(d-3)(d-4)$ which must obey $4\alpha/L^{2}\leq 1$ to avoid naked singularity, $M$ is a constant of integration which
is related to the black hole horizon through $M=\frac{(d-2)\Sigma r_{+}^{d-1}}{16\pi GL^{2}}$  with $\Sigma$ the volume of the
($d-3$)-dimensional Ricci flat space.
 Note that in the asymptotic region
($r\rightarrow\infty$), we have
\begin{eqnarray}
f(r)\sim\frac{r^2}{2\alpha}\left[ 1-\sqrt{1-\frac{4\alpha}{L^2}} \right]\,.
\end{eqnarray}
We can define the effective asymptotic AdS scale by \cite{Gregory}
\begin{eqnarray}\label{LeffAdS}
L^2_{\rm eff}=\frac{2\alpha}{1-\sqrt{1-\frac{4\alpha}{L^2}}} \to
\left\{
\begin{array}{rl}
L^2   \ , &  \quad {\rm for} \ \alpha \rightarrow 0~, \\
\frac{L^2}{2}  \ , &  \quad {\rm for} \  \alpha \rightarrow
\frac{L^2}{4}~.
\end{array}\right.
\end{eqnarray}
We note that the limit $L^{2}=4 \alpha$ is known as the Chern-Simons limit~\cite{Crisostomo:2000bb}. The Hawking temperature of the black hole,
which will be interpreted as the temperature of the CFT, can be easily obtained
\begin{eqnarray}
\label{Hawking temperature} T=\frac{(d-1)r_{+}}{4\pi L^{2}}\ .
\end{eqnarray}

In the background of $d$-dimensional Gauss-Bonnet-AdS black hole,
we consider a Maxwell field and a charged complex scalar field
with the action
\begin{eqnarray}\label{System}
S=\int d^{d}x\sqrt{-g}\left[
-\frac{1}{4}F_{\mu\nu}F^{\mu\nu}-|\nabla\psi - iA\psi|^{2}
-m^2|\psi|^2 \right] \ .
\end{eqnarray}
We assume that these fields are weakly coupled to gravity, so they do not backreact on the metric (probe limit). Taking the ansatz $\psi=|\psi|$,
$A=\phi dt$ where $\psi$, $\phi$ are both functions of $r$ only, we can obtain the equations of motion for $\psi, \phi$
\begin{eqnarray}
\psi^{\prime\prime}+\left(
\frac{f^\prime}{f}+\frac{d-2}{r}\right)\psi^\prime
+\left(\frac{\phi^2}{f^2}-\frac{m^2}{f}\right)\psi=0\,, \label{Psi}
\end{eqnarray}
\begin{eqnarray}
\phi^{\prime\prime}+\frac{d-2}{r}\phi^\prime-\frac{2\psi^2}{f}\phi=0~.
\label{Phi}
\end{eqnarray}
It was argued that the coupling of the scalar field to the Maxwell field can produce a negative effective mass \cite{GubserPRD78} which will
become more important at low temperature leading to an instability of the $\psi=0$ configuration resulting in the black hole to acquire
hair~\cite{HartnollPRL101,HartnollJHEP12}.

The equations (8) and (9) can be solved numerically by doing integration from the horizon out to the infinity. The regularity condition at the
horizon gives the boundary conditions $\psi(r_{+})=f^\prime(r_{+})\psi^\prime(r_{+})/m^{2}$ and $\phi(r_{+})=0$. At the asymptotic region
($r\rightarrow\infty$), the solutions behave like
\begin{eqnarray}
\psi=\frac{\psi_{-}}{r^{\lambda_{-}}}+\frac{\psi_{+}}{r^{\lambda_{+}}}\,,\hspace{0.5cm}
\phi=\mu-\frac{\rho}{r^{d-3}}\,, \label{infinity}
\end{eqnarray}
with
\begin{eqnarray}
\lambda_\pm=\frac{1}{2}\Big{[}(d-1)\pm\sqrt{(d-1)^{2}+4m^{2}L_{\rm
eff}^2}~\Big{]}\,,\label{LambdaZF}
\end{eqnarray}
where $\mu$ and $\rho$ are interpreted as the chemical potential and charge density in the dual field theory respectively. The coefficients
$\psi_{-}$ and $\psi_{+}$ both multiply normalizable modes of the scalar field equations and according to the AdS/CFT correspondence, they
correspond to the vacuum expectation values $\psi_{-}=<\mathcal{O}_{-}>$, $\psi_{+}=<\mathcal{O}_{+}>$ of an operator $\mathcal{O}$ dual to the
scalar field. We can impose boundary conditions that either $\psi_{-}$ or $\psi_{+}$ vanish. As was noted in~\cite{HartnollPRL101} imposing
boundary conditions in which both $\psi_{-}$ and $\psi_{+}$ are non-zero makes the asymptotic AdS theory
unstable~\cite{Hertog:2004rz,Hertog:2005hu}.

\section{Scalar condensation in the Gauss-Bonnet-AdS black hole
background}

In the case of flat Schwarzschild-AdS black hole background, the scalar operators $\langle{\cal O}_{-}\rangle$ and $\langle{\cal O}_{+}\rangle$
have different behaviors at low temperatures~\cite{HartnollPRL101}. While the condensate $\langle{\cal O}_{+}\rangle$ have similar behavior to
the BCS theory, the condensate $\langle{\cal O}_{-}\rangle$ diverges at low temperatures. As it was shown in~\cite{HartnollJHEP12} this result
was obtained because the backreaction on the metric was neglected. In this section we will present a detailed analysis of the condensation of
these operators which are subjected to curvature corrections.

\subsection{The condensation for the scalar operator $\langle{\cal O}_{+}\rangle$ }

A study of the condensation of the scalar operator $\langle{\cal O}_{+}\rangle$ in five-dimensional Gauss-Bonnet-AdS black hole background was
carried out in \cite{Gregory}. It was found that for fixed mass of the scalar field to a value of $m^2=-3/L^2$, the increase of the Gauss-Bonnet
coupling $\alpha$ results in the decrease of the critical temperature so that the higher curvature corrections make it harder for the scalar
field to condense. This can be understood invoking the arguments presented in \cite{GubserPRD78}. The scalar hair can be formed just outside the
horizon because the electromagnetic repulsion of the charged scalar field can overcome the gravitational attraction, resulting in the
condensation of the scalar field bouncing off the AdS boundary. In our case because of the strong curvature effects outside the horizon this
mechanism is less effective. The maximum effect is obtained in the Chern-Simons limit where the Gauss-Bonnet theory is strongly coupled
\cite{Charmousis:2008kc}.

\begin{figure}[ht]
\includegraphics[scale=0.75]{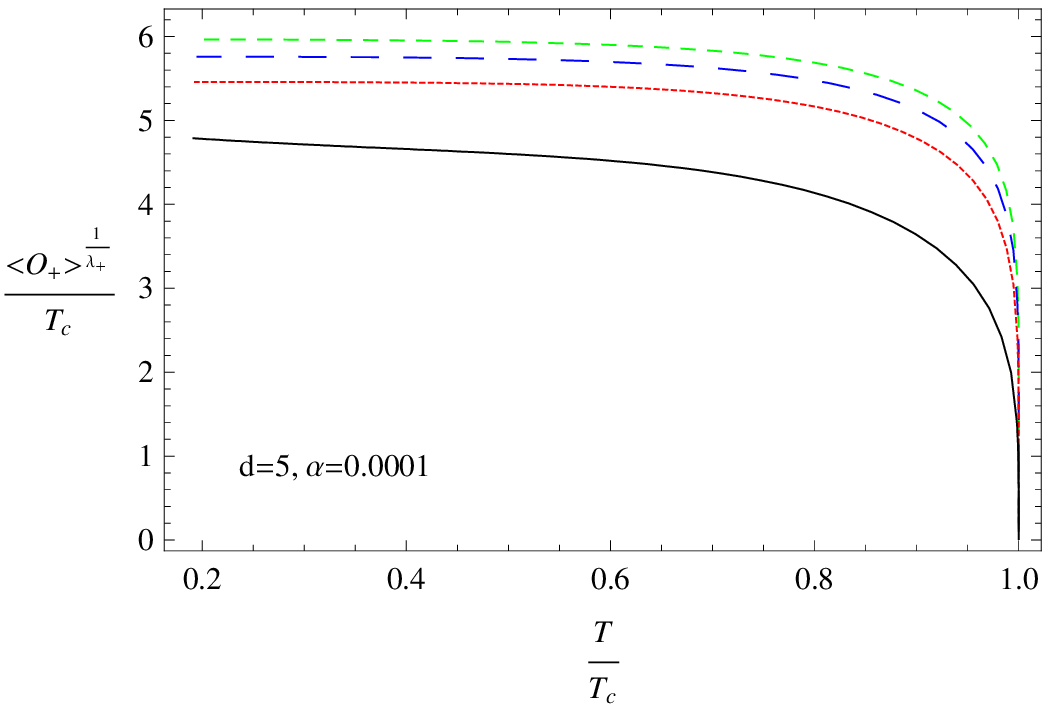}\hspace{0.2cm}%
\includegraphics[scale=0.75]{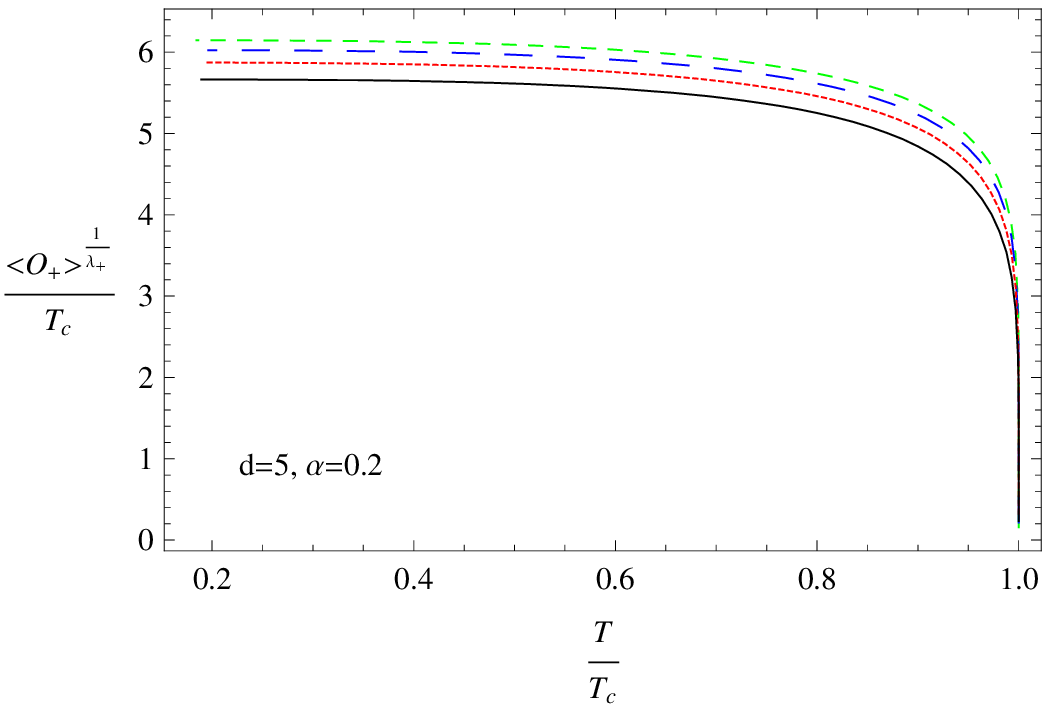}\\ \vspace{0.0cm}
\caption{\label{Cond-d5mass} (color online) The condensate as a function of temperature with fixed values $\alpha=0.0001$ and $0.2$ for various
masses of the scalar field in $d=5$ dimension. The four lines from bottom to top correspond to increasing mass, i.e., $m^2L^2=-4$ (black), $-3$
(red), $-2$ (blue) and $-1$ (green) respectively.}
\end{figure}

Changing the mass of the scalar field, we present in Fig.1 the
scalar mass influence on the condensation. It is clear that for the
same $\alpha$, the condensation gap becomes larger if $m^2$ becomes
less negative. The difference caused by the influence of the scalar
mass will become smaller when  there is higher curvature correction
in the AdS background.

\begin{figure}[H]
\includegraphics[scale=0.75]{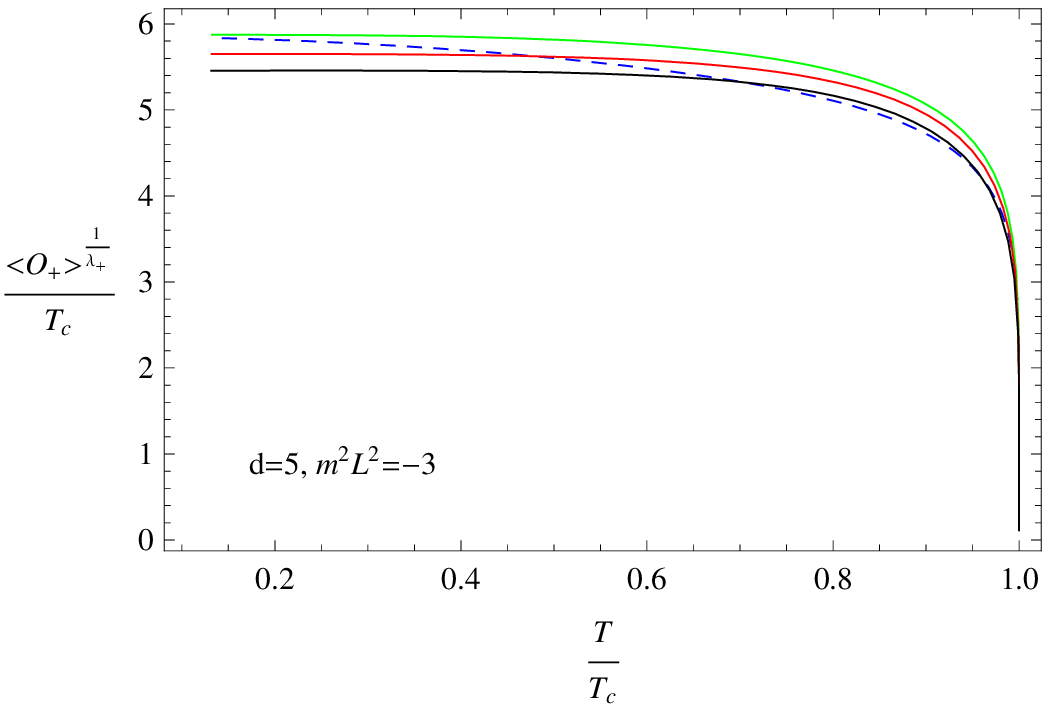}\hspace{0.2cm}%
\includegraphics[scale=0.75]{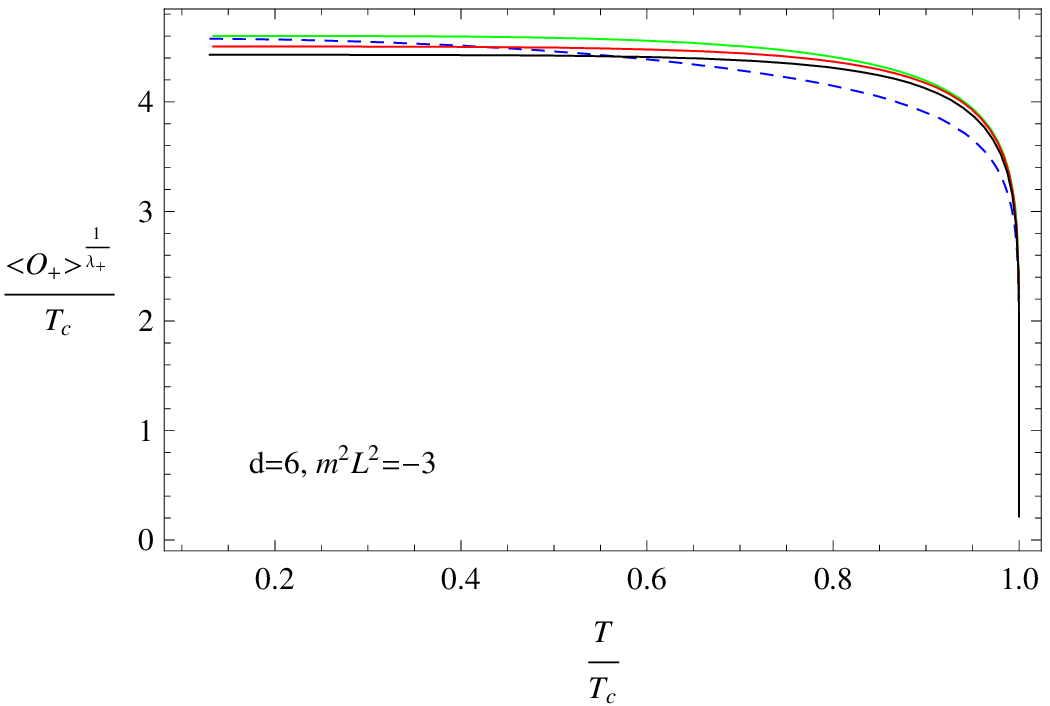}\\ \vspace{0.0cm}
\caption{\label{Cond-D5D6} (color online) The condensate as a function of temperature with different values of $\alpha$ for fixed mass of the
scalar field $m^2L^{2}=-3$ in $d=5$ and $6$ dimensions. The three lines from bottom to top correspond to increasing $\alpha$, i.e.,
$\alpha=0.0001$ (black), $0.1$ (red) and $0.2$ (green) respectively. And the other blue dashed line corresponds to $\alpha=0.25$.}
\end{figure}

In Fig.2 we further show the influence the dimensionality of the
spacetime has on the scalar condensation. For choosing the same
value of the scalar mass,  qualitative features occur as we vary
$\alpha$  in  AdS black hole background of different
dimensionality. However, as the spacetime dimension increases, the
condensation gap becomes smaller for the same $\alpha$, which
means that the scalar hair can be formed easier in the
higher-dimensional background. Moreover we observed that the
difference caused by the curvature corrections are reduced when
the spacetime dimension becomes higher.

\subsubsection{Analytical understanding of the condensation}

Since equations (8) and (9) are coupled and nonlinear, we have to rely on the numerical calculation. A semi-analytical method can be applied in
understanding the condensation. The method was first applied in the calculation of the Grey-Body factors of black holes~\cite{Kanti:2002ge}. The
method consists in finding approximate solutions near the horizon and in the asymptotic (A)dS space and then smoothly match the solutions at an
intermediate point. In particular in \cite{Gregory} an analytic expression for the critical temperature was obtained, matching smoothly  the
leading order solutions near the horizon and asymptotically at an intermediate point and the phase transition phenomenon was demonstrated. It was
further shown that the critical temperature obtained in this way is in a good agreement with the numerical result. It was argued that the
qualitative features of the approximation do not change with the arbitrary choice of the intermediate matching point.

In this subsection we extend the analytic approach of
\cite{Gregory} to d-dimensional AdS black holes with Gauss-Bonnet
terms. Rewriting Eqs. (8) and (9) with a new coordinate
$z=r_{+}/r$, we have
\begin{eqnarray}
&&\psi^{\prime\prime}+\left(\frac{f^\prime}{f}-\frac{d-4}{z}\right)\psi^\prime
+\frac{r_{+}^2}{z^4}\left(\frac{\phi^2}{f^2}-\frac{m^{2}}{f}\right)
\psi=0~, \label{PsiZ}\\
&&\phi^{\prime\prime}-\frac{d-4}{z}\phi^\prime
-\frac{r_{+}^{2}}{z^4}\frac{2\psi^2}{f}\phi=0~, \label{PhiZ}
\end{eqnarray}
where the prime denotes differentiation in $z$. Regularity at the
horizon $z=1$ requires
\begin{eqnarray}
\psi(1)=-\frac{d-1}{m^{2}L^{2}}\psi^\prime(1)\,,\hspace{0.5cm}\phi(1)=0\,. \label{horizon regularity}
\end{eqnarray}
Near the AdS boundary $z=0$,
\begin{eqnarray}
\psi=C_{-}z^{\lambda_{-}}+C_{+}z^{\lambda_{+}}\,, \hspace{0.5cm}\phi=\mu-\frac{\rho}{r_{+}^{d-3}}z^{d-3}\,. \label{boundary behavior}
\end{eqnarray}
We will  set $C_{-}=0$ and fix $\rho$ in the following discussion.

Expanding Eqs. (\ref{PsiZ}) and (\ref{PhiZ}) near $z=1$ with the regular horizon boundary condition (\ref{horizon regularity}), one can easily
obtain the leading order approximate solutions near the horizon
\begin{eqnarray}
\psi(z)&=&\left(1+\frac{m^{2}L^{2}}{d-1}\right)\psi(1)-\frac{m^{2}L^{2}}{d-1}\psi(1)z
\nonumber\\
&&+\frac{1}{4}\left\{\left[2+\frac{m^{2}L^{2}}{d-1}-\frac{2(d-1)\alpha}{L^{2}}\right]
\frac{m^{2}L^{2}}{d-1}-\frac{L^{4}}{(d-1)^{2}r_{+}^{2}}\phi^{\prime}(1)^2\right\}
\psi(1)(1-z)^2\cdots, \label{psiTaylor}
\end{eqnarray}
\begin{eqnarray}
\phi(z)=-\phi^\prime(1)(1-z)+\frac{1}{2}\left[(d-4)-\frac{2L^2}{d-1}\psi(1)^2\right] \phi^{\prime}(1)(1-z)^2+\cdots. \label{phiTaylor}
\end{eqnarray}
The solutions near the asymptotic AdS region can be read off from
Eq. (\ref{boundary behavior}),
\begin{eqnarray}
\psi=C_{+}z^{\lambda_{+}}\,, \hspace{0.5cm}\phi=\mu-\frac{\rho}{r_{+}^{d-3}}z^{d-3}\,. \label{boundary Taylor}
\end{eqnarray}

To match smoothly the solutions (\ref{psiTaylor}), (\ref{phiTaylor}) and (\ref{boundary Taylor})  at an intermediate point $z_m$ with $0<z_m<1$,
we can use the following equations which connect the above two asymptotic regions
\begin{eqnarray}
C_{+}z_{m}^{\lambda_+}&=&\left(1+\frac{m^{2}L^{2}}{d-1}\right)a-\frac{m^{2}L^{2}}{d-1}z_{m}a
\nonumber\\
&&+\frac{1}{4}\left\{\left[2+\frac{m^{2}L^{2}}{d-1}-\frac{2(d-1)\alpha}{L^{2}}\right]
\frac{m^{2}L^{2}}{d-1}-\frac{L^{4}}{(d-1)^{2}r_{+}^{2}}b^2\right\}(1-z_{m})^2a~,
\label{match psi}
\end{eqnarray}
\begin{eqnarray}
\lambda_{+}C_{+}z_{m}^{\lambda_{+}-1}&=&-\frac{m^{2}L^{2}}{d-1}a
\nonumber
\\
&-&\frac{1}{2}\left\{\left[2+\frac{m^{2}L^{2}}{d-1}-\frac{2(d-1)\alpha}{L^{2}}\right]
\frac{m^{2}L^{2}}{d-1}-\frac{L^{4}}{(d-1)^{2}r_{+}^{2}}b^2\right\}(1-z_{m})a
\label{match dpsi}
\end{eqnarray}
\begin{eqnarray}
&&\mu-\frac{\rho}{r_{+}^{d-3}}z_{m}^{d-3}=(1-z_{m})b-
\frac{1}{2}\left[(d-4)-\frac{2L^2}{d-1}a^2\right](1-z_{m})^{2}b~,
\label{match phi}
\end{eqnarray}
\begin{eqnarray}
&&-(d-3)\frac{\rho}{r_{+}^{d-3}}z_{m}^{d-4}=
-b+\left[(d-4)-\frac{2L^2}{d-1}a^2\right](1-z_{m})b~, \label{match
dphi}
\end{eqnarray}
where $\psi(1)\equiv a$ and $-\phi^\prime(1)\equiv b$ with $a,b>0$ which makes $\psi(z)$ and $\phi(z)$ positive near the horizon. Using Eqs.
(\ref{match psi}) and (\ref{match dpsi}), we can eliminate $b$ and get
\begin{eqnarray}
C_+=\frac{2(d-1)+m^{2}L^{2}(1-z_{m})}{(d-1) [2z_{m}+(1-z_{m})\lambda_+]z_{m}^{\lambda_{+}-1}}\; a. \label{C psi}
\end{eqnarray}
Substituting (\ref{C psi}) into (\ref{match dpsi}), we find
\begin{eqnarray}
b&=&2(d-1)\frac{r_{+}}{L^2}\times \\ \nonumber
&&\sqrt{\frac{[2(d-1)+m^{2}L^2(1-z_{m})]\lambda_+}
{2(d-1)(1-z_{m})[2z_{m}+(1-z_{m})\lambda_+]}+\frac{(2-z_{m})m^{2}L^2}{2(d-1)(1-z_{m})}
+\left[\frac{m^{2}L^2}{2(d-1)}\right]^{2}-\frac{m^{2}\alpha}{2}}~.
\label{b phi}
\end{eqnarray}
Similarly, from Eqs. (\ref{match phi}) and (\ref{match dphi}) we can express $a$ as
\begin{eqnarray}
a^2=\frac{(d-1)(d-3)z_{m}^{d-4}\rho}{2(1-z_{m})L^2r_{+}^{d-3}b} \left\{1-\frac{[1+(4-d)(1-z_{m})]r_{+}^{d-3}b}{(d-3)z_{m}^{d-4}\rho}\right\}.
\label{a psi}
\end{eqnarray}
Using Eq. (\ref{b phi}) and the Hawking temperature (\ref{Hawking temperature}), we can rewrite (\ref{a psi}) as
\begin{eqnarray}
a^2=\frac{(d-1)[1+(4-d)(1-z_{m})]}{2(1-z_{m})L^2}\left(\frac{T_c}{T}\right)^{d-2} \left[1-\left(\frac{T}{T_c}\right)^{d-2}\right], \label{rewrite
a psi}
\end{eqnarray}
and the critical temperature $T_c$ is given by
\begin{eqnarray}
T_c=\frac{d-1}{4\pi L^{2}}\left\{\frac{(d-3)z_{m}^{d-4}L^{2}\rho}{[1+(4-d)(1-z_{m})]\tilde{b}}\right\}^{\frac{1}{d-2}}, \label{match Tc}
\end{eqnarray}
where we have set $b=\tilde{b}r_{+}/L^{2}$.

Following the AdS/CFT dictionary, we obtain the relation
\begin{eqnarray}
\langle {\cal O_{+}} \rangle\equiv L C_{+} r_{+}^{\lambda_+} L^{-2\lambda_{+}}=L C_{+}\left(\frac{4\pi T}{d-1}\right)^{\lambda_+}.
\end{eqnarray}
Thus, from (\ref{C psi}) and (\ref{rewrite a psi}) the expectation value $\langle {\cal O_{+}} \rangle$ is given by
\begin{eqnarray}
\label{operator} \frac{\langle {\cal O_{+}} \rangle^{\frac{1}{\lambda_+}}}{T_c}=\Upsilon~\frac{T}{T_{c}} \left\{\left(\frac{T_c}{T}\right)^{d-2}
\left[1-\left(\frac{T}{T_c}\right)^{d-2}\right]\right\}^{\frac{1}{2\lambda_{+}}},
\end{eqnarray}
where $\Upsilon$ is defined by
\begin{eqnarray}
\Upsilon=\frac{4\pi}{d-1}\left\{\frac{\sqrt{(d-1)[1+(4-d)(1-z_{m})]} ~[2(d-1)+m^{2}L^{2}(1-z_{m})]}{\sqrt{2(1-z_{m})}(d-1)
[2z_{m}+(1-z_{m})\lambda_+]z_{m}^{\lambda_{+}-1}}\right\}^{\frac{1}{\lambda_{+}}}.
\end{eqnarray}

It is interesting to observe that using  Eq. (\ref{b phi}) and  Eq. (\ref{match Tc}) we find that for
\begin{eqnarray}
z_{md}=\frac{d-5}{d-4}~, \label{divergence point}
\end{eqnarray}
the critical temperature $T_c$ diverges! Therefore, if we use the value $z_{m}=1/2$ in $d=6$ the method breaks down, contrary to what it was
found in \cite{Gregory} for $d=5$. This shows that the matching point is not truly arbitrary. In order to get the correct critical temperature
$T_c$ for $d>5$, we have to choose the matching point $z_{m}$ in the range $z_{md}<z_{m}<1$.

In table \ref{Tc-D6} we present the critical temperature obtained
analytically by fixing $z_{m}=7/10$ for $d=6$ and its comparison
with numerical results. We observe that when the mass of the
scalar field is nonzero, selecting the matching point from the
appropriate range, we can obtain consistent analytic result with
that obtained numerically.  When the scalar mass is zero, the
analytic approximation fails to give the correct critical
temperature dependence on the Gauss-Bonnet term. This can also be
seen in d=5. The reason is that in the analytic approximation, the
Gauss-Bonnet term is entangled with the mass of the scalar field
as shown in (\ref{b phi}). Therefore setting $m=0$ the $\alpha$
contribution in the analytic result is eliminated.

\begin{table}[ht]
\begin{center}
\caption{\label{Tc-D6} The critical temperature $T_{c}$  obtained
by the analytical method (left column) and the numerical method
(right column) for $d=6$. The matching point is set as
$z_{m}=7/10$ which satisfies the range $\frac{d-5}{d-4}<z_{m}<1$.
We have used $\rho=1$ in the table.}
\begin{tabular}{c c c c c}
         \hline
$\alpha$ &0.0001&0.1&0.2& 0.25
        \\
        \hline
$m^2L^2=0$&$0.249$~~~~$0.249$&$0.249$~~~~$0.242$&
 $0.249$~~~~$0.232$&$0.249$~~~~$0.223$
          \\
$m^2L^2=-1$&$0.253$~~~~$0.253$&$0.253$~~~~$0.246$&
 $0.252$~~~~$0.235$&$0.252$~~~~$0.225$
          \\
$m^2L^2=-2$&$0.258$~~~~$0.257$&$0.257$~~~~$0.250$&
 $0.256$~~~~$0.238$&$0.255$~~~~$0.227$
          \\
$m^2L^2=-3$&$0.264$~~~~$0.264$&$0.262$~~~~$0.255$&
 $0.260$~~~~$0.242$&$0.259$~~~~$0.229$
          \\
$m^2L^2=-4$&$0.271$~~~~$0.271$&$0.268$~~~~$0.261$&
 $0.265$~~~~$0.246$&$0.263$~~~~$0.232$
          \\
        \hline
\end{tabular}
\end{center}
\end{table}

In summary, we have reexamined the analytic approach in understanding the condensation. Although the position of the matching point does not
change the qualitative result when $d=5$, it cannot be arbitrary when spacetime dimension is higher. To avoid a breakdown of the method, the
matching point has to be in an appropriate range of values. When the scalar mass is nonzero, the critical temperature obtained by analytic method
agrees well with that calculated numerically. When the scalar mass is zero, the
 Gauss-Bonnet term does not contribute to  the analytic
approximation, so that we cannot count on the analytic method as it was also observed in \cite{Gregory}.

\subsubsection{Conductivity}

It was argued in \cite{HorowitzPRD78} that in (2+1) and (3+1)-dimensional superconductors a universal relation connecting the gap frequency in
conductivity with the critical temperature $T_c$ holds $\omega_g/T_c\approx 8$, to better than 10 $\%$ for a range of scalar masses. This is
roughly twice the BCS value 3.5 indicating that the holographic superconductors are strongly coupled. However it was  found in \cite{Gregory}
that this relation is not stable in the presence of the Gauss-Bonnet correction terms. In this section we will further examine this result.

In the Gauss-Bonnet black hole background, the Maxwell equation at zero spatial momentum and with a time dependence of the form $e^{-iwt}$ gives
\begin{eqnarray}
A_{x}^{\prime\prime}+\left(\frac{f^\prime}{f}+\frac{d-4}{r}\right)A_{x}^\prime +\left(\frac{\omega^2}{f^2}-\frac{2\psi^2}{f}\right)A_{x}=0 \; ,
\label{Maxwell Equation}
\end{eqnarray}
where we used the ansatz for the perturbed Maxwell field $\delta A_{x}=A_{x}(r)e^{-i\omega t}dx$. To avoid the complicated behavior in the gauge
field falloff in dimensions higher than five, we restrict our study to $d=5$.  We solve the above equation with an ingoing wave boundary
condition $A_{x}(r)\sim f(r)^{-\frac{i\omega}{4r_+}}$ near the horizon.  The general behavior in the asymptotic AdS region ($r\rightarrow\infty$)
is seen to be
\begin{eqnarray}\label{Maxwell boundary}
A_{x}=A^{(0)}+\frac{A^{(2)}}{r^2} +\frac{A^{(0)}\omega^2 L_{\rm
eff}^2}{2} \frac{\log\Lambda r}{r^2}~.
\end{eqnarray}
Removing the divergence by an appropriate boundary counter term,
we have the conductivity obtained in  \cite{HorowitzPRD78}
\begin{eqnarray}\label{GBConductivity}
\sigma=\frac{2A^{(2)}}{i\omega A^{(0)}}+\frac{i\omega}{2} \ .
\end{eqnarray}

\begin{figure}[H]
\includegraphics[scale=0.375]{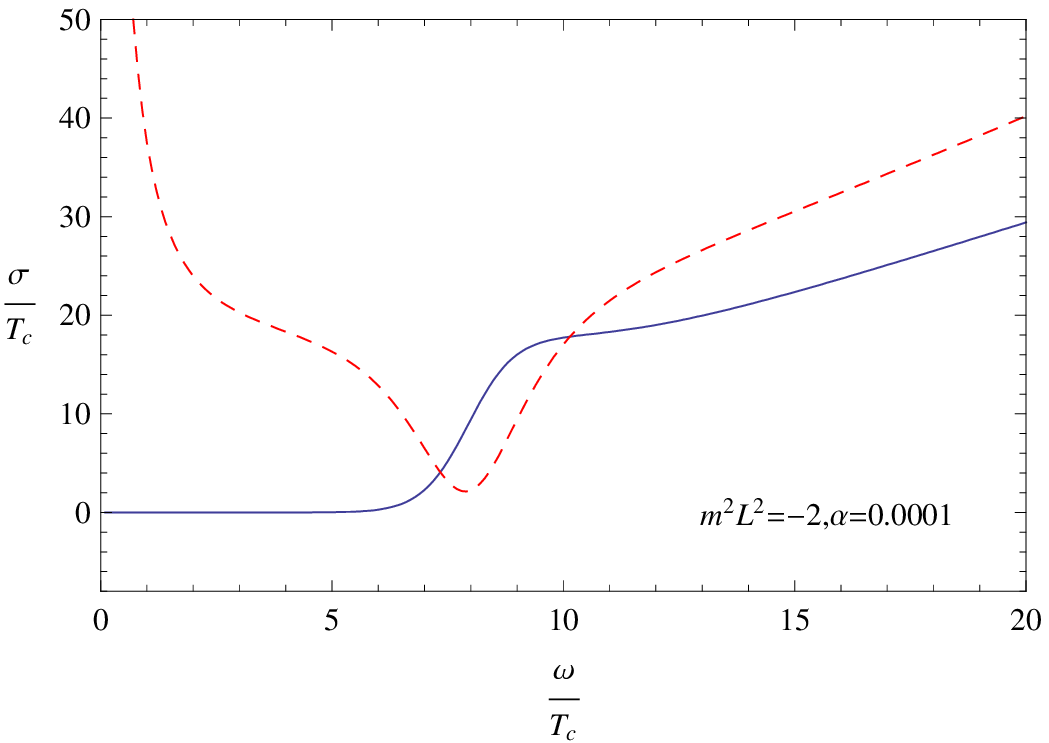}\hspace{0.2cm}%
\includegraphics[scale=0.375]{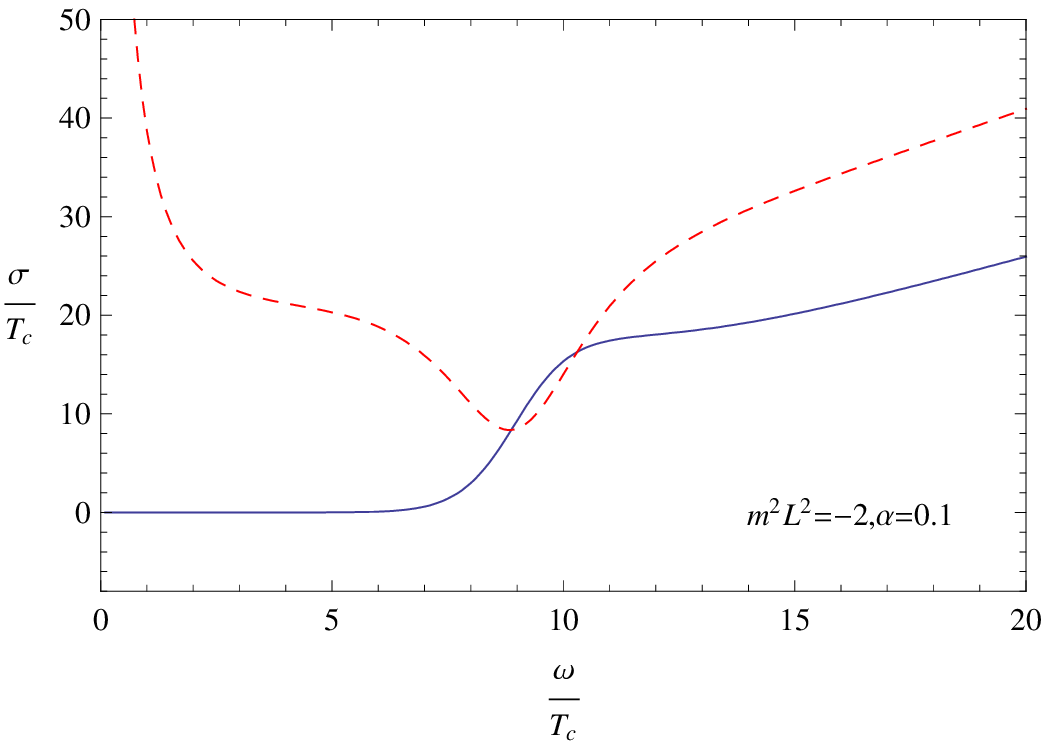}\hspace{0.2cm}%
\includegraphics[scale=0.375]{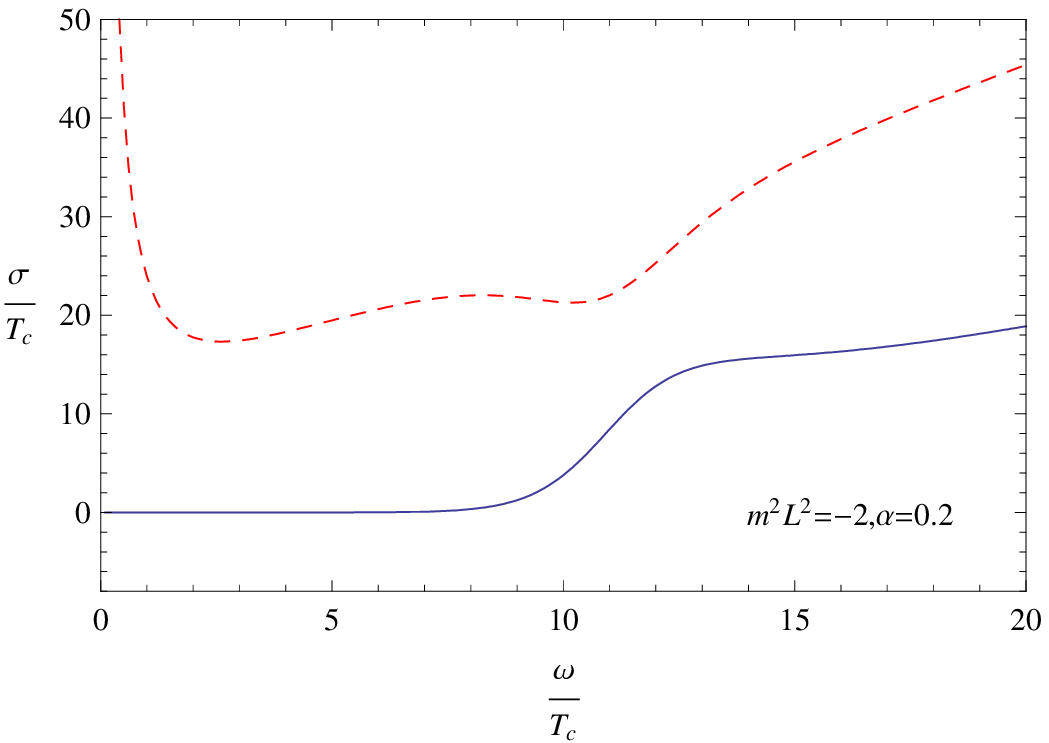}\hspace{0.2cm}%
\includegraphics[scale=0.375]{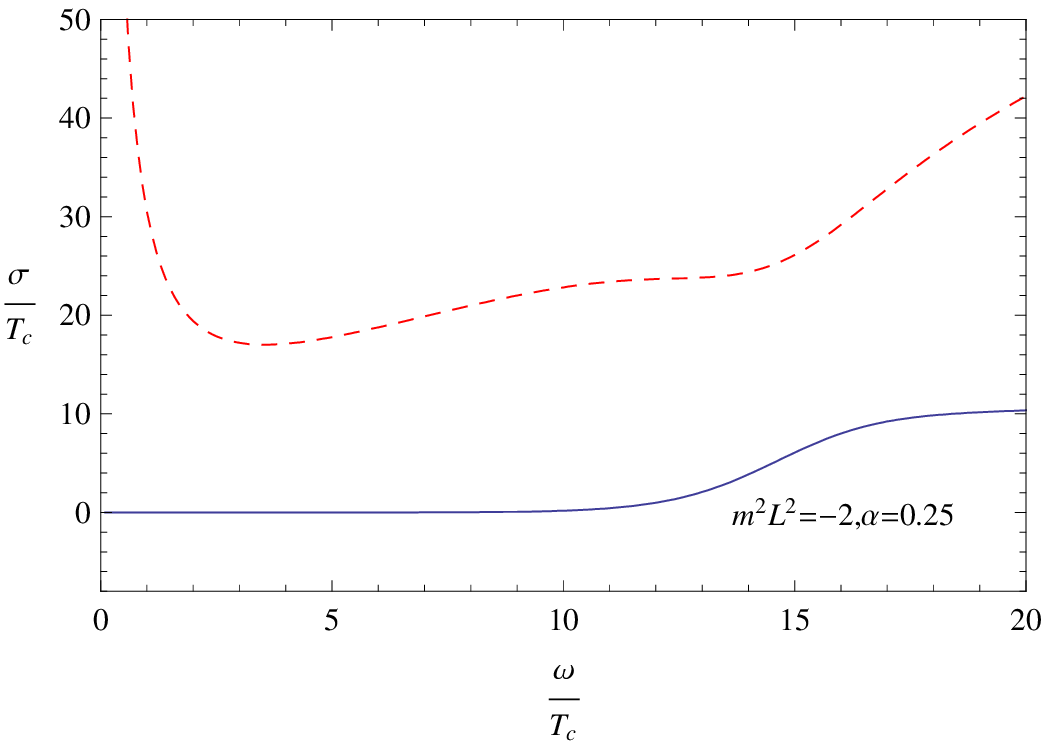}\\ \vspace{0.0cm}
\includegraphics[scale=0.375]{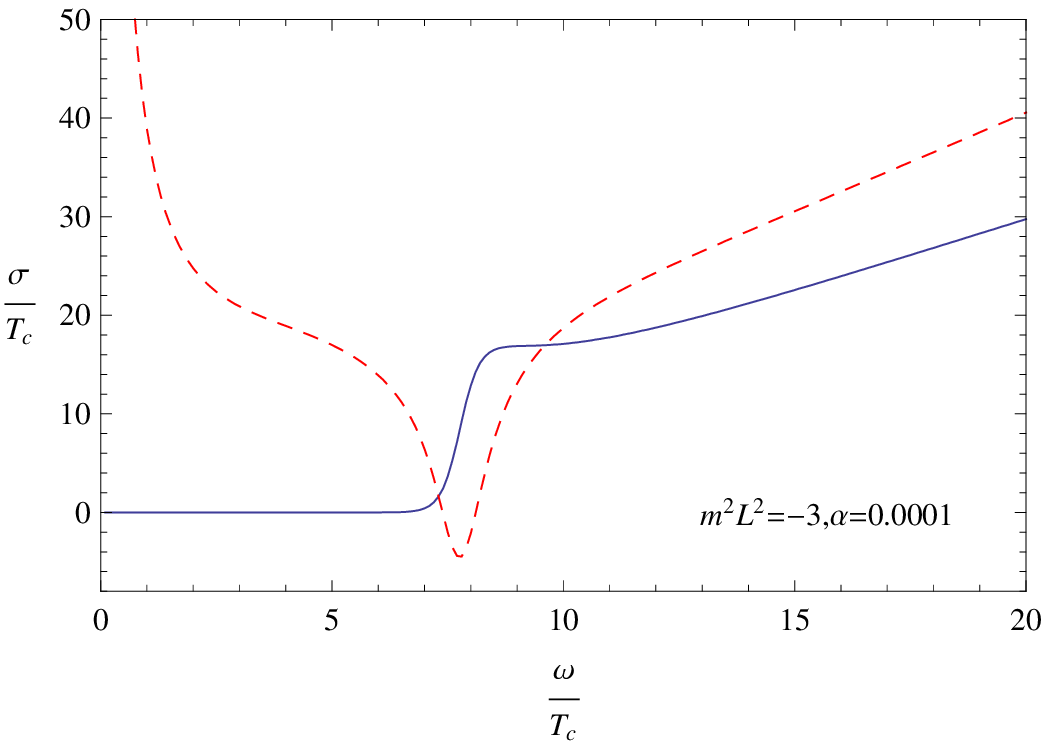}\hspace{0.2cm}%
\includegraphics[scale=0.375]{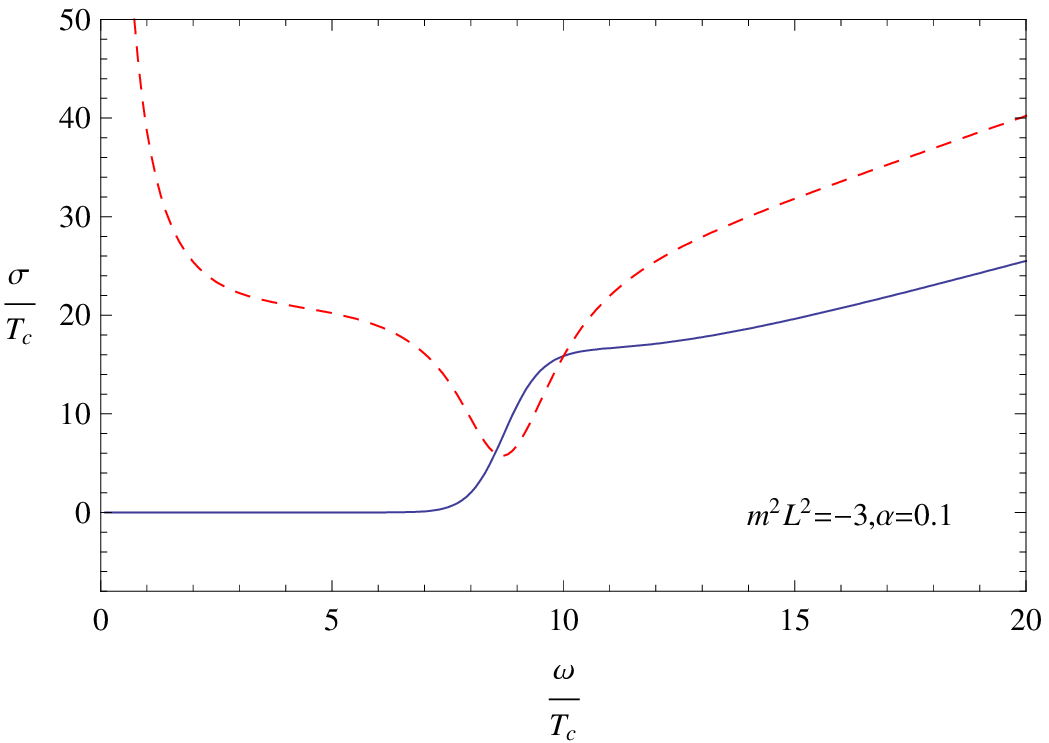}\hspace{0.2cm}%
\includegraphics[scale=0.375]{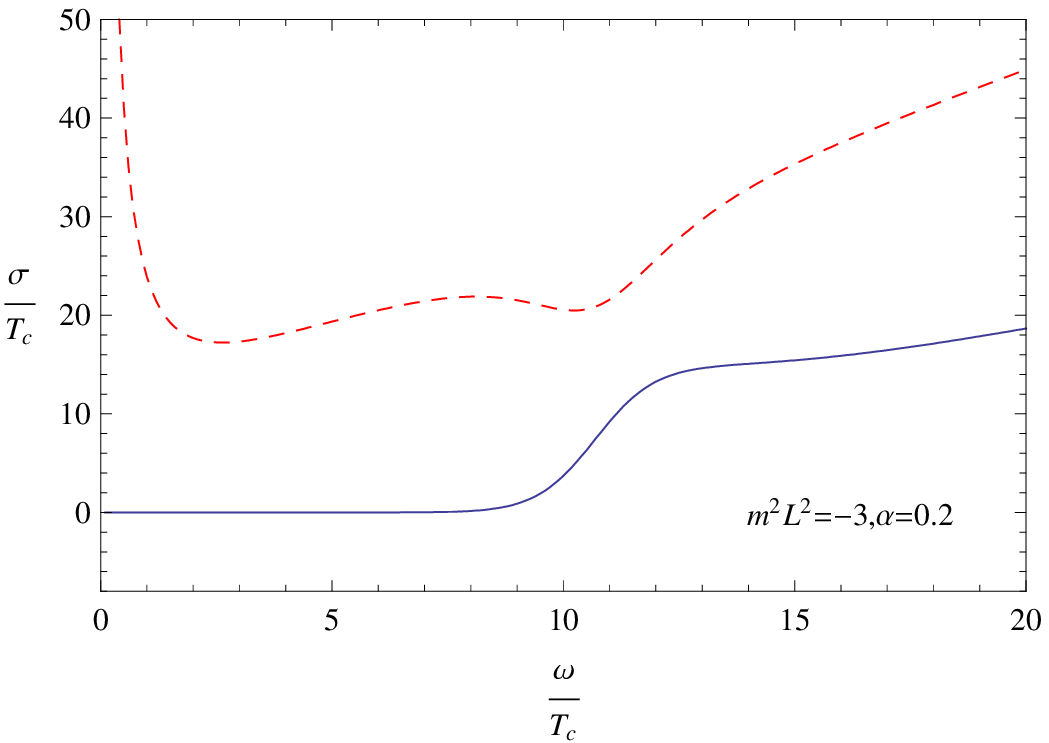}\hspace{0.2cm}%
\includegraphics[scale=0.375]{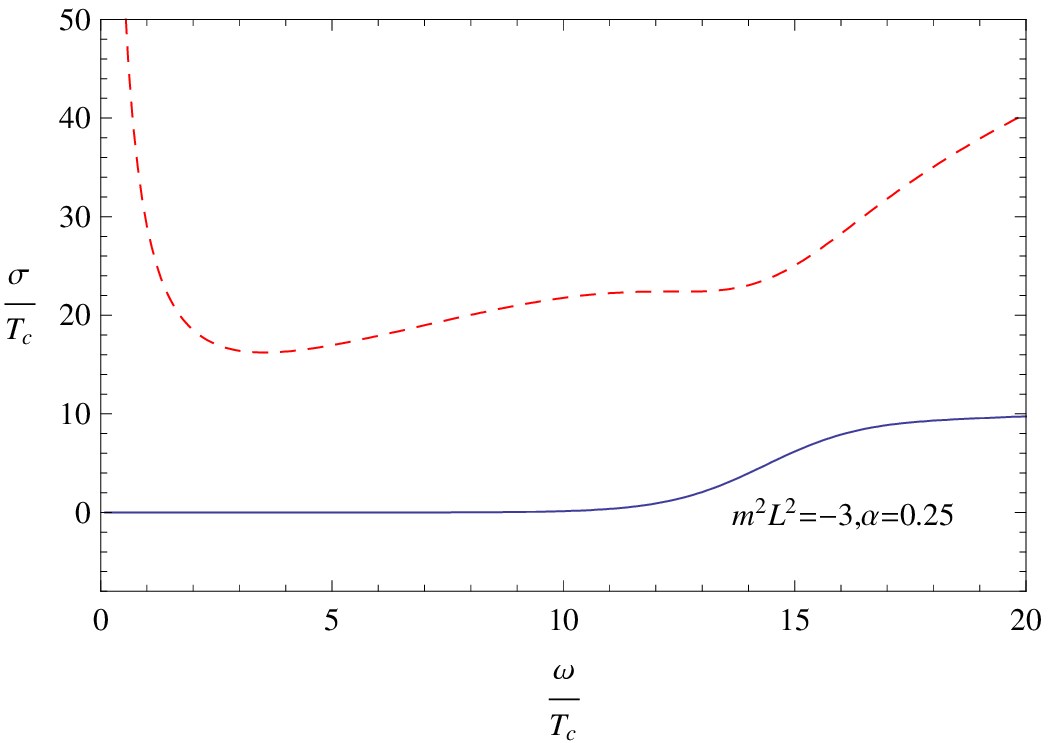}\\ \vspace{0.0cm}
\includegraphics[scale=0.375]{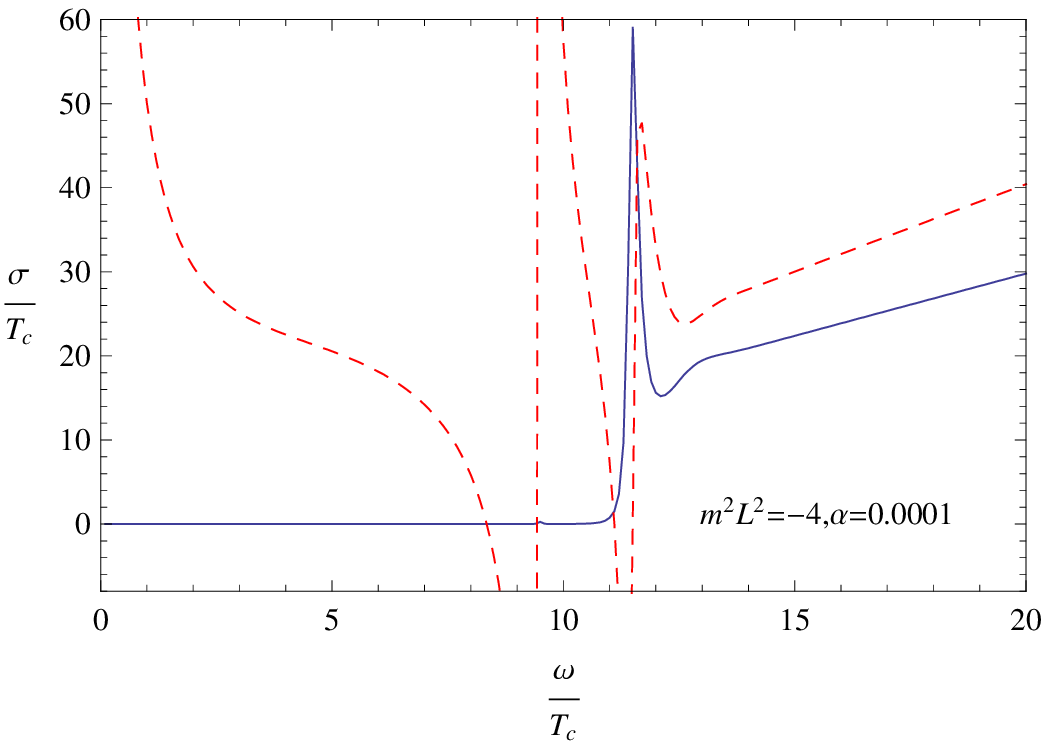}\hspace{0.2cm}%
\includegraphics[scale=0.375]{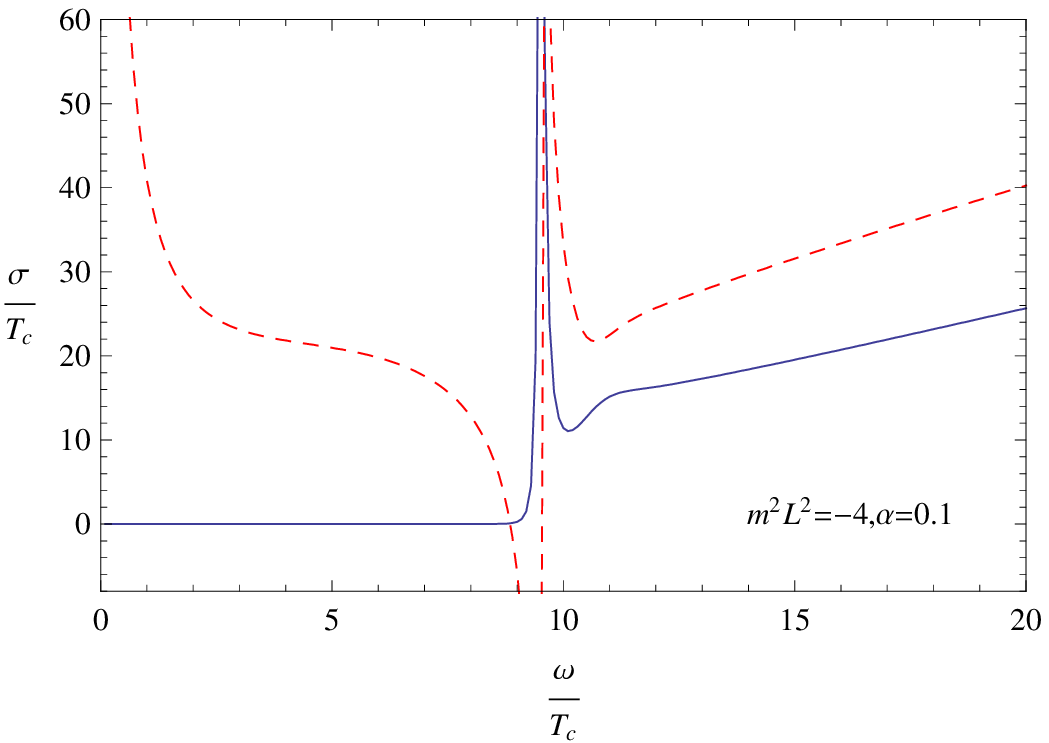}\hspace{0.2cm}%
\includegraphics[scale=0.375]{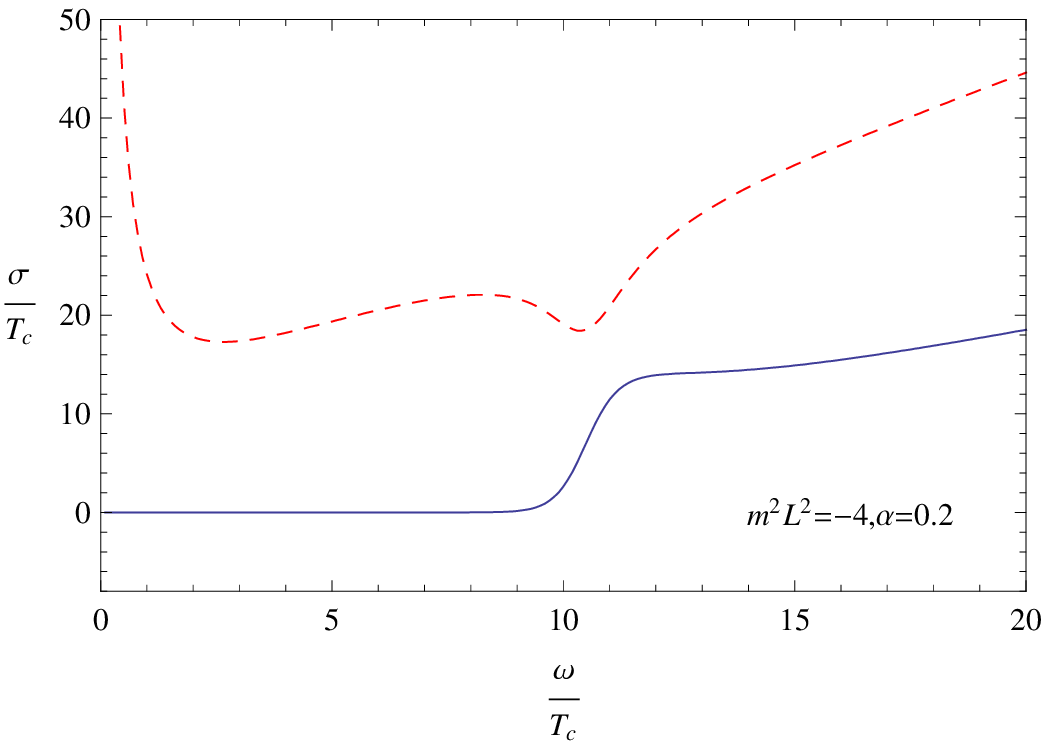}\hspace{0.2cm}%
\includegraphics[scale=0.375]{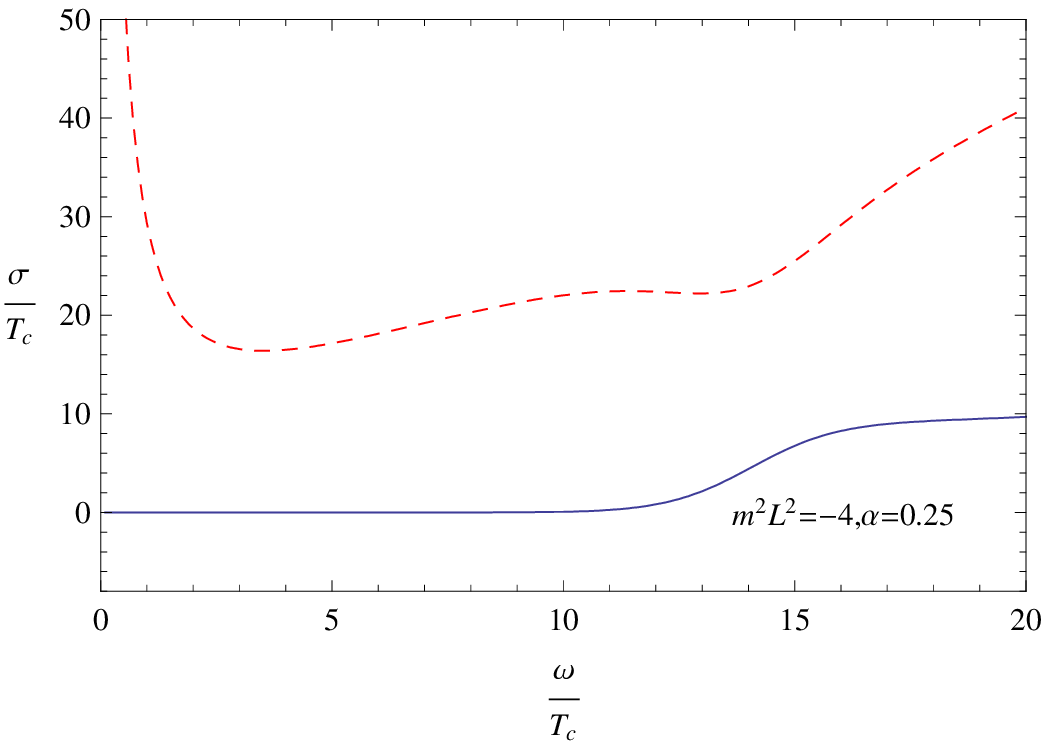}\\ \vspace{0.0cm}
\caption{\label{Conductivity} (color online) Conductivity for
($3+1$)-dimensional Gauss-Bonnet superconductors with fixed values
of $\alpha$ for different mass of the scalar field.}
\end{figure}

\begin{table}[ht]
\begin{center}
\caption{\label{ConductivityTc} The ratio $\omega_{g}/T_{c}$ with
fixed values of $\alpha$ for different mass of the scalar field. }
\begin{tabular}{|c|c|c|c|c|}
         \hline
$~~\alpha~~$ &~~0.0001~~&~~0.1~~&~~0.2~~&~~0.25~~
          \\
        \hline
~~$m^{2}L^{2}=-2$~~ & ~~$7.8$~~ & ~~$8.8$~~ & ~~$10.4$~~& ~~$14.2$~~
          \\
        \hline
$m^{2}L^{2}=-3$ & $7.7$ & $8.6$ & $10.2$ & $14.1$
          \\
        \hline
$m^{2}L^{2}=-4$ & ~~~ & ~~~ & $10.2$ & $13.7$
          \\
         \hline
\end{tabular}
\end{center}
\end{table}

In Fig. \ref{Conductivity} we plot the frequency dependent conductivity obtained by solving the Maxwell equation numerically for $\alpha=0.0001,
0.1$, $0.2$ and $0.25$ with different mass of the scalar field, i.e., $m^2L^2=-2$, $-3$ and $-4$. The blue (solid) line represents the real part,
and red (dashed) line is the imaginary part of $\sigma$. We find a gap in the conductivity with the gap frequency $\omega_{g}$. For the same mass
of the scalar field, we observe that with the increase of the Gauss-Bonnet coupling constant, the gap frequency $\omega_{g}$ becomes larger.

We observe in table \ref{ConductivityTc} that for increasing Gauss-Bonnet coupling we have larger deviations from the value $\omega_g/T_c\approx
8$ with the maximum value attained in the Chern-Simons limit. This shows that the high curvature corrections really changes the expected
universal relation in the gap frequency. On the other hand, if we concentrate on the same Gauss-Bonnet coupling, we observed that the change of
the mass of the scalar field has little effect on the gap frequency as was also observed in \cite{HorowitzPRD78}.

\subsection{The condensation for the scalar operator $\langle{\cal O}_{-}\rangle$}

In this section we will impose the condition $\psi_{+}=0$ and study the condensation generated by the scalar operator $\langle{\cal
O}_{-}\rangle$. As we already learned, in the case of four-dimensional Schwarzschild-AdS black hole the condensation of this scalar operator for
small temperatures diverges \cite{HorowitzPRD78}. We expect the same behavior to occur also in the presence of curvature corrections. Thus, if we
want to keep the probe limit approximation the temperatures considered should not be very small.

The presence of the Gauss-Bonnet correction term gives the
possibility of choosing the scalar mass as $m^2 L_{\rm eff}^2$.
This choice is closely related to $\lambda$ in Eq.
(\ref{LambdaZF}), which is the dimension of the boundary operator
dual to the scalar field. On the other hand, fixing the scalar
mass by $m^{2}L_{\rm eff}^2$ contains the influence of
Gauss-Bonnet coupling  by considering Eq. (\ref{LeffAdS}).
Therefore, it is more appropriate to choose the scalar mass by the
value of $m^{2}L_{\rm eff}^2$ when  the Gauss-Bonnet correction
terms are included. However, this choice of the scalar mass should
be in the range $-\frac{(d-1)^{2}}{4}<m^{2}L_{\rm
eff}^2<-\frac{(d-1)^{2}}{4}+1$ where both modes of the asymptotic
values of the scalar fields are normalizable
\cite{HorowitzPRD78,Umeh}, except at the saturation of the BF
bound $-(d-1)^{2}/4$ \cite{Breitenloher}.

We could have used the scalar mass as $m^{2}L_{\rm eff}^2$ in
studying the condensation of the operator $\langle{\cal
O}_{+}\rangle$ in the previous section. However, as it was shown
in \cite{Gregory}  the qualitative features in condensation are
the same in varying $\alpha$ if the mass of the scalar field is
chosen as $m^2 L^2$ or alternatively $m^2 L_{\rm eff}^2$.

\begin{figure}[H]
\includegraphics[scale=0.75]{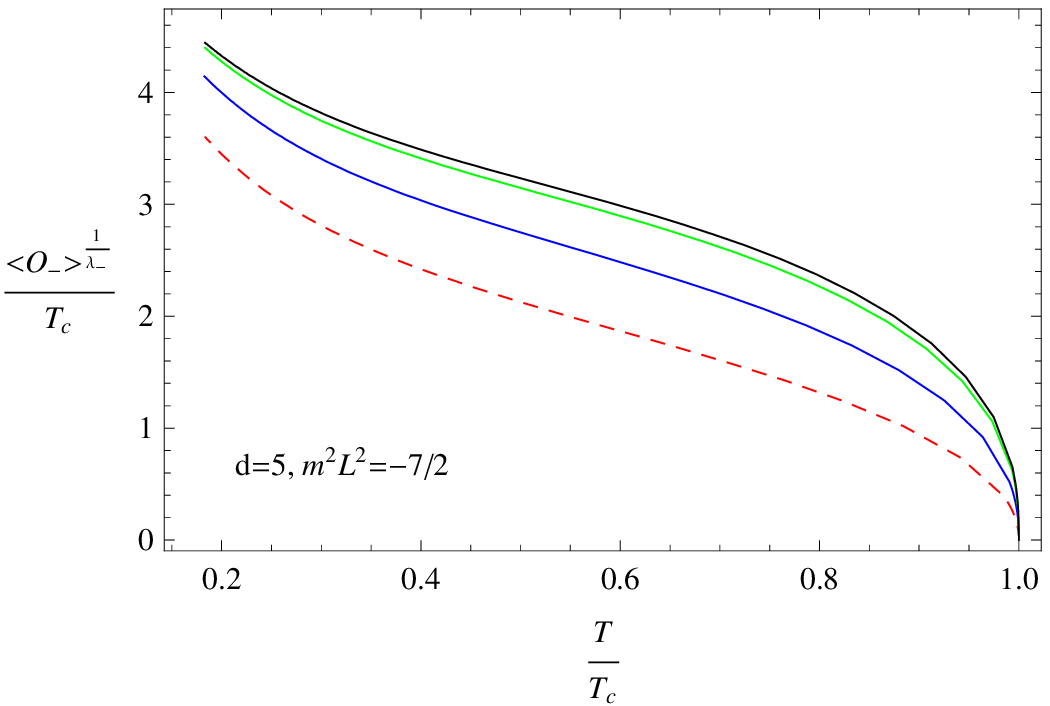}\hspace{0.2cm}%
\includegraphics[scale=0.75]{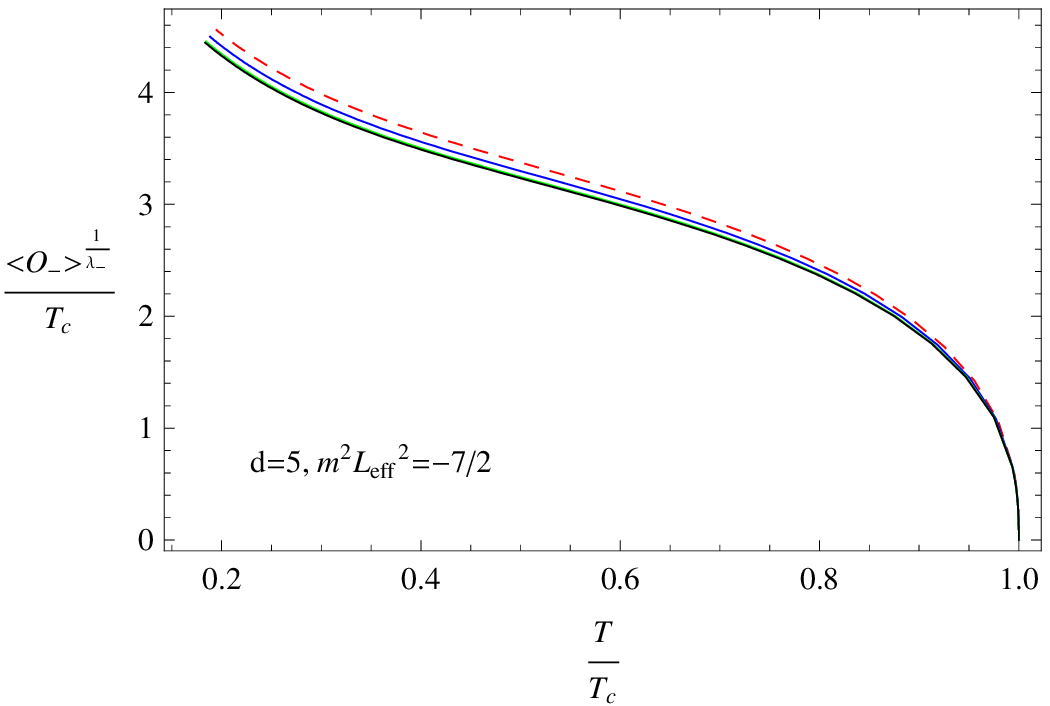}\\ \vspace{0.0cm}
\caption{\label{Cond-mass-LambdaF} (color online) The condensate
as a function of temperature with different values of $\alpha$ if
we fix the mass of the scalar field by $m^2L^{2}$ (the left panel)
or by $m^{2}L_{\rm eff}^2$ (the right panel) for $\psi_{-}$ when
$\psi_{+}$ vanishes. In the left panel four lines from top to
bottom correspond to increasing $\alpha$, i.e., $\alpha=0.0001$
(black), $0.01$ (green), $0.05$ (blue) and $0.1$ (red and dashed)
respectively, but they are arranged the other way round in the
right panel.}
\end{figure}

Nevertheless,  as shown in Fig. \ref{Cond-mass-LambdaF}, the
condensation for the scalar operator $\langle{\cal O}_{-}\rangle$
for different choices of the mass of the scalar field, have
completely different behavior as $\alpha$ is changing. Selecting the
value of $m^{2}L_{\rm eff}^2$ for the scalar mass we get the same
qualitative dependence of the condensation as in the case of the
condensation  for the scalar operator $\langle{\cal O}_{+}\rangle$
(right panel of Fig. \ref{Cond-mass-LambdaF}), while choosing the
value of $m^2L^2$ for the scalar mass we get the opposite behavior
(left panel of Fig. \ref{Cond-mass-LambdaF}).

\begin{figure}[ht]
\includegraphics[scale=0.75]{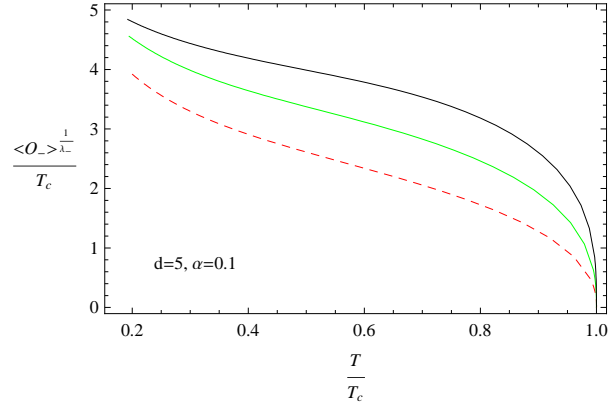}\vspace{0.0cm}
\caption{\label{CondBHD5Fa01} (color online) The condensate as a
function of temperature with different values of $m^{2}L_{\rm
eff}^2$ with $\alpha=0.1$ for $\psi_{-}$ when $\psi_{+}$ vanishes.
The three lines from top to bottom correspond to increasing mass
of the scalar field, i.e., $m^{2}L_{\rm eff}^2=-15/4$ (black),
$-7/2$ (green) and $-13/4$ (red and dashed) respectively.}
\end{figure}

Fixing the Gauss-Bonnet coupling $\alpha$, in Fig.
\ref{CondBHD5Fa01} we show the dependence of the condensation of the
scalar operator $\langle{\cal O}_{-}\rangle$ on the mass of scalar
filed. It is observed that it is qualitative different than the
behavior of the condensation of  the scalar operator $\langle{\cal
O}_{+}\rangle$: the larger mass of the scalar field makes it easier
for the scalar hair  to form. This is consistent with the result
found in \cite{HorowitzPRD78}.

\section{Holographic Superconductor in Gauss-Bonnet-AdS soliton}

 In this section we will study a holographic dual of a Gauss-Bonnet-AdS soliton. Motivated by the work presented in
\cite{Nishioka-Ryu-Takayanagi} we will extend their discussion to the Ricci flat AdS soliton in the Gauss-Bonnet gravity and examine the effect
of the Gauss-Bonnet term on the condensation and conductivity.

\subsection{Scalar condensation in the AdS soliton}

By analytically continuing the Ricci flat black hole
  one obtains the AdS
soliton in the Gauss-Bonnet gravity \cite{Cai-Kim-Wang}
\begin{eqnarray}\label{soliton}
ds^2=-r^{2}dt^{2}+\frac{dr^2}{f(r)}+f(r)d\varphi^2+r^{2}h_{ij}dx^{i}dx^{j}.
\end{eqnarray}
Obviously, there does not exist any horizon but a conical singularity at $r=r_{s}$ in this solution. Imposing a period $\beta=\frac{4\pi
L^{2}}{(d-1)r_{s}}$ for the coordinate $\varphi$, we can remove the singularity.

Beginning with the Einstein-Maxwell-scalar theory (\ref{System}), we can get the equations of motion for the scalar field $\psi$ and gauge field
$\phi$ in the form
\begin{eqnarray}
\psi^{\prime\prime}+\left(
\frac{f^\prime}{f}+\frac{d-2}{r}\right)\psi^\prime
+\left(\frac{\phi^2}{r^2f}-\frac{m^2}{f}\right)\psi=0\,, \label{Psi}
\end{eqnarray}
\begin{eqnarray}
\phi^{\prime\prime}+\left(\frac{f^\prime}{f}+\frac{d-4}{r}\right)
\phi^\prime-\frac{2\psi^2}{f}\phi=0.
\label{Phi}
\end{eqnarray}
We will solve these two equations numerically with appropriate boundary conditions at $r=r_{s}$ and at the boundary $r\rightarrow\infty$. The
solutions near the AdS boundary are the same as Eq. (\ref{infinity}). At the tip $r=r_{s}$, the solutions behave as
\begin{eqnarray}
\psi=\tilde{\psi}_{0}+\tilde{\psi}_{1}\log(r-r_{s})+\tilde{\psi}_{2}(r-r_{s})+\cdots\,, \nonumber \\
\phi=\tilde{\phi}_{0}+\tilde{\phi}_{1}\log(r-r_{s})+\tilde{\phi}_{2}(r-r_{s})+\cdots\,, \label{SolitonBoundary}
\end{eqnarray}
where $\tilde{\psi}_{i}$ and $\tilde{\phi}_{i}$ ($i=0,1,2,\cdots$)
 are the integration constants. In order to keep every physical quantity
finite, we impose the Neumann-like boundary condition $\tilde{\psi}_{1}=\tilde{\phi}_{1}=0$ \cite{Nishioka-Ryu-Takayanagi} in our discussion.
Obviously, one can find a constant nonzero gauge field $\phi(r_{s})$ at $r=r_{s}$, in contrary to that of the AdS black hole where
$\phi(r_{+})=0$ at the horizon. We will still use the probe approximation in our calculation and select the mass of the scalar field in the range
$-\frac{(d-1)^{2}}{4}<m^{2}L_{\rm eff}^2<-\frac{(d-1)^{2}}{4}+1$ where both modes of the asymptotic values of the scalar fields are normalizable.
For clarity, we will take $d=5$  and one can easily extend the study to $d\geq6$.

\begin{figure}[H]
\includegraphics[scale=0.75]{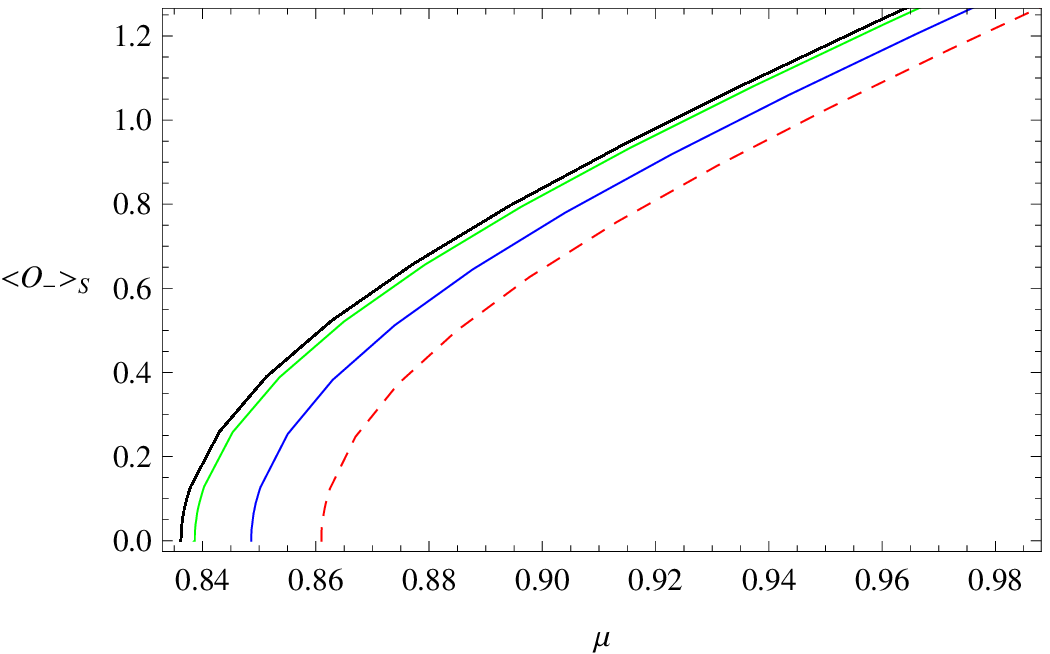}\hspace{0.2cm}%
\includegraphics[scale=0.75]{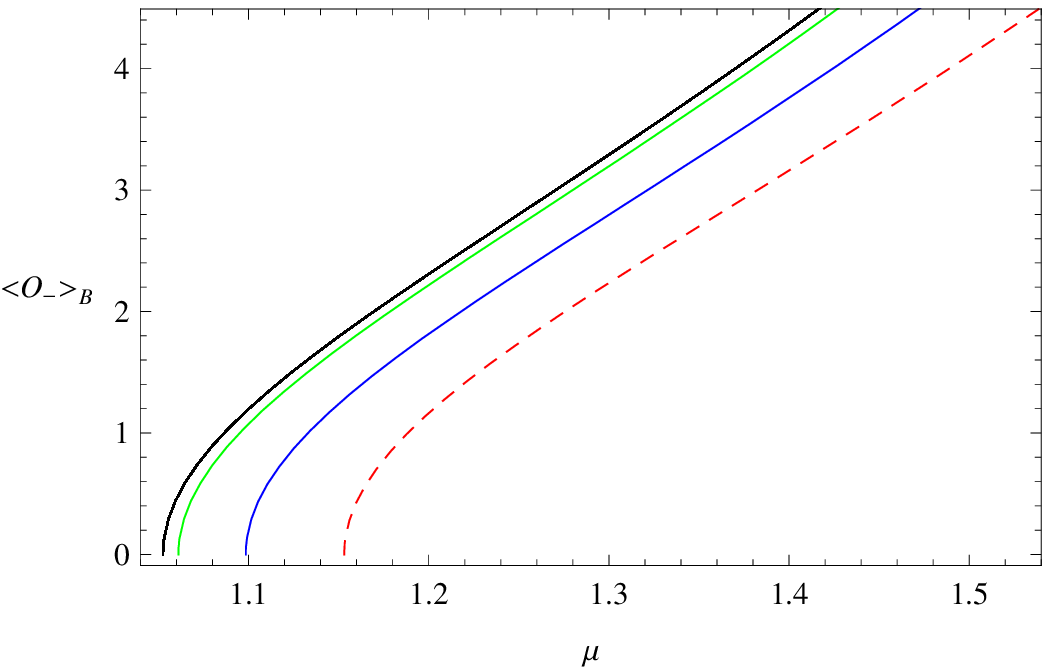}\\ \vspace{0.0cm}
\includegraphics[scale=0.75]{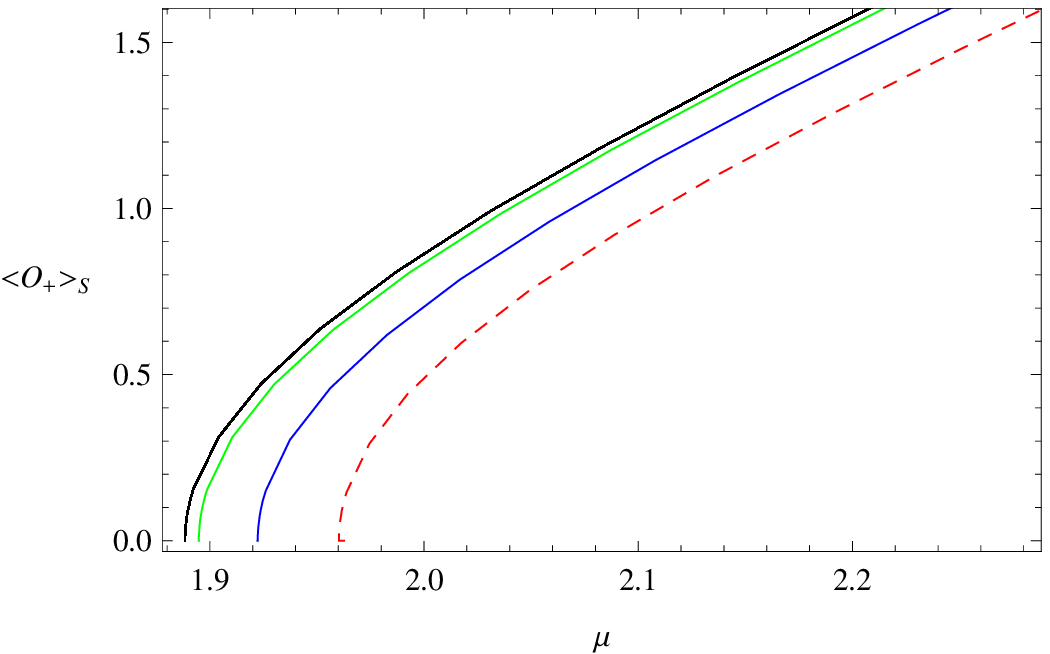}\hspace{0.2cm}%
\includegraphics[scale=0.75]{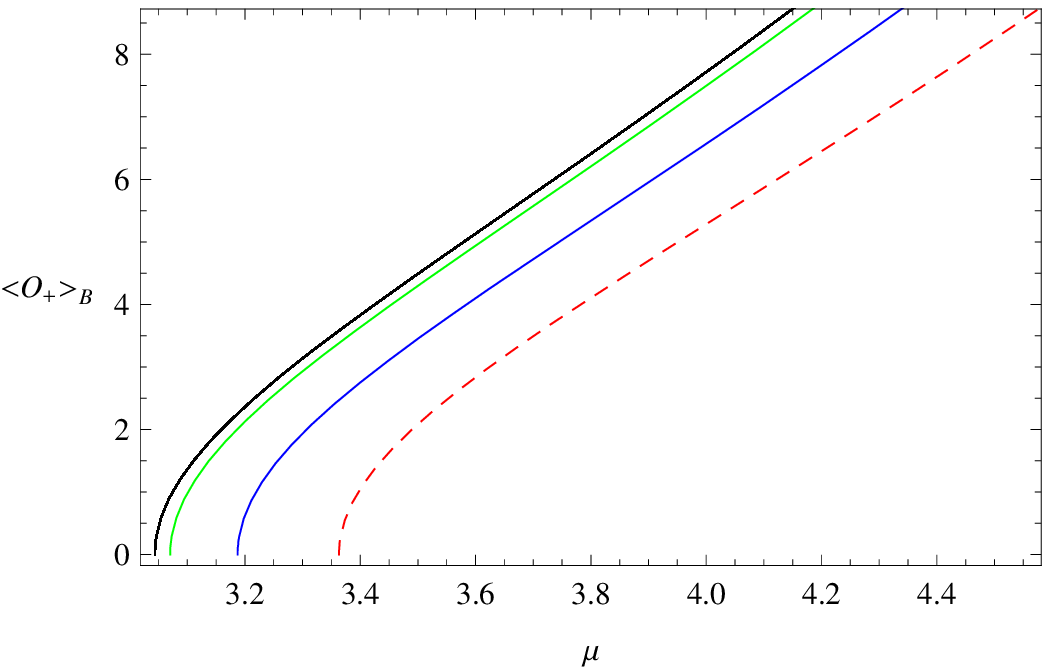}\\ \vspace{0.0cm}
\caption{\label{Cond-SolitonBH} (color online) The condensates of
the scalar operators $\langle{\cal O_{-}}\rangle$ and $\langle{\cal
O_{+}}\rangle$ with respect to the chemical potential $\mu$ in the
Gauss-Bonnet Soliton (left column) and Gauss-Bonnet Black hole
(right column). We fix the mass of the scalar field by $m^{2}L_{\rm
eff}^2=-15/4$ and the four lines from left to right correspond to
increasing $\alpha$, i.e., $\alpha=0.0001$ (black), $0.01$ (green),
$0.05$ (blue) and $0.1$ (red and dashed) respectively.}
\end{figure}

In Fig. \ref{Cond-SolitonBH} we plot the condensations of scalar operators $\langle{\cal O_{-}}\rangle_{S}$ and $\langle{\cal O_{+}}\rangle_{S}$
with respect to the chemical potential $\mu$ in the Gauss-Bonnet-AdS Soliton for different Gauss-Bonnet coupling constants with the fixed scalar
mass $m^{2}L_{\rm eff}^2=-15/4$ (two panels of the left column). The condensation occurs for scalar operators $\langle{\cal O}_{i}\rangle_{S}$
($i=\pm$) with different values of $\alpha$ if $\mu>\mu_{i S}$, where $\mu_{i S}$ is the so-called critical chemical potential for scalar
operators $\langle{\cal O}_{i}\rangle_{S}$ which just begin to condense. Thus, we obtain $\mu_{- S}$ and $\mu_{+ S}$ for scalar operators
$\langle{\cal O_{-}}\rangle_{S}$ and $\langle{\cal O_{+}}\rangle_{S}$ with different values of $\alpha$ respectively, i.e., $\mu_{-S}=0.836$ and
$\mu_{+S}=1.888$ for $\alpha=0.0001$; $\mu_{-S}=0.839$ and $\mu_{+S}=1.895$ for $\alpha=0.01$; $\mu_{-S}=0.849$ and $\mu_{+S}=1.922$ for
$\alpha=0.05$ and $\mu_{-S}=0.861$ and $\mu_{+S}=1.960$ for $\alpha=0.1$. For the same mass of the scalar field, it is observed that the critical
chemical potential $\mu_{\pm S}$ for the condensation to occur increases as $\alpha$ increases. This result is consistent with that observed in
the AdS black hole as shown in the right column in Fig. \ref{Cond-SolitonBH}. For the AdS soliton, we again find that the higher curvature
correction makes it harder for the scalar hair to form which is similar to that seen in the AdS black hole. For the same $\alpha$,
$\mu_{-S}<\mu_{+S}$, which agrees to that found in \cite{Nishioka-Ryu-Takayanagi} implying that it is easier for the scalar condensation to be
formed in the scalar operators $\langle{\cal O_{-}}\rangle_{S}$.

\begin{figure}[H]
\includegraphics[scale=0.75]{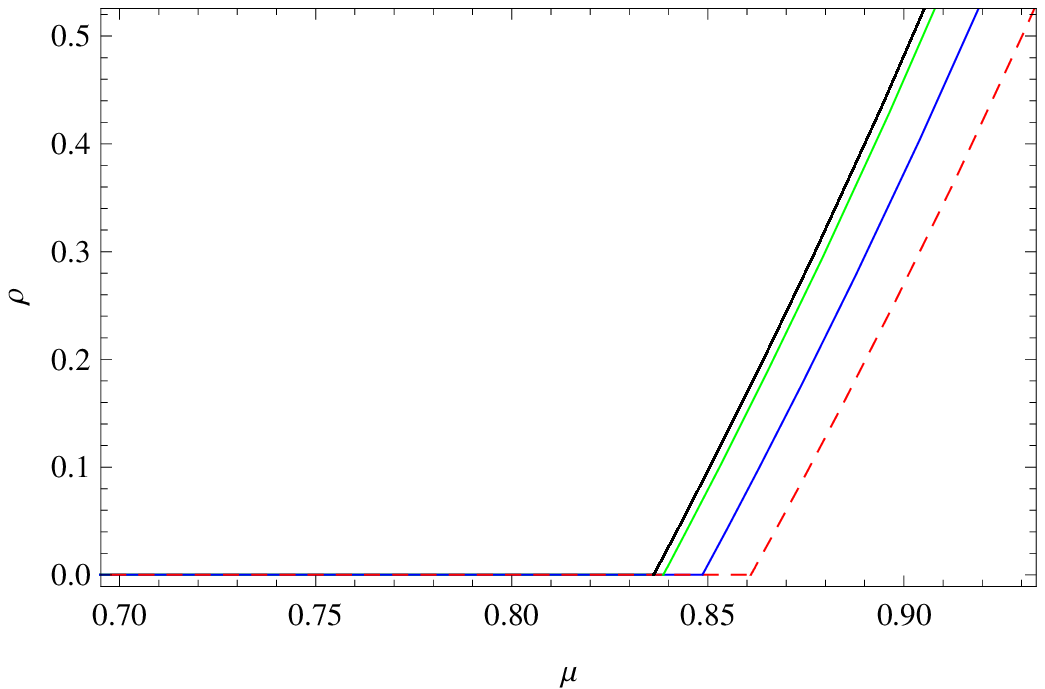}\hspace{0.2cm}%
\includegraphics[scale=0.75]{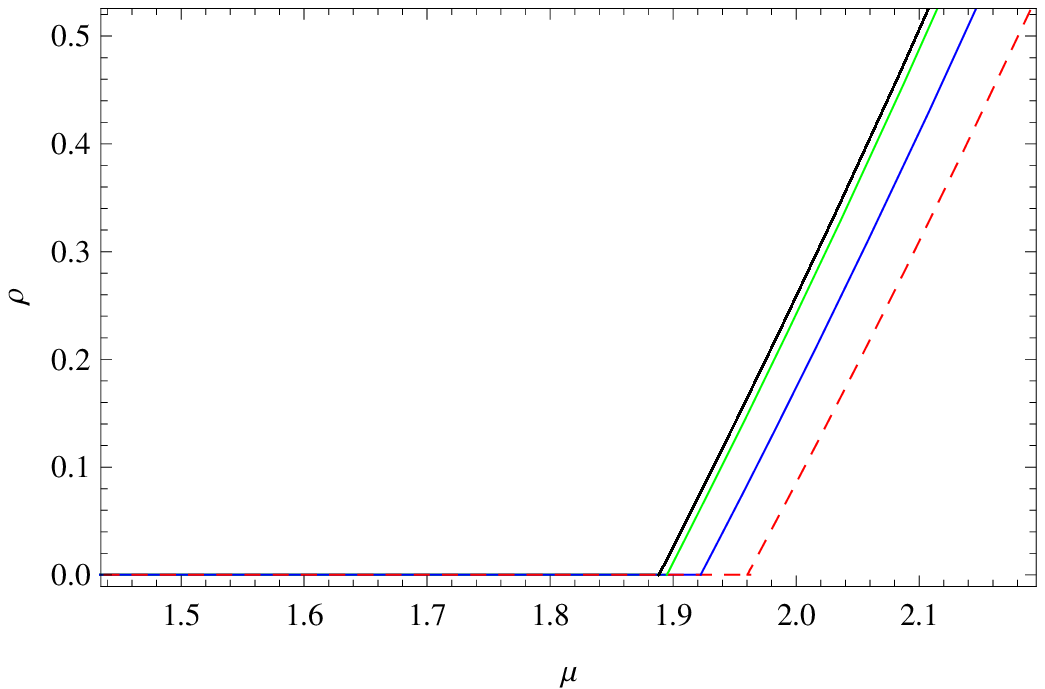}\\ \vspace{0.0cm}
\caption{\label{SLRealization} (color online) The charge density
$\rho$ as a function of the chemical potential $\mu$ with different
values of $\alpha$ when $\langle{\cal O_{-}}\rangle\neq0$ (left) and
$\langle{\cal O_{+}}\rangle\neq0$ (right). We fix the mass of the
scalar field by $m^{2}L_{\rm eff}^2=-15/4$ and the four lines from
left to right correspond to increasing $\alpha$, i.e.,
$\alpha=0.0001$ (black), $0.01$ (green), $0.05$ (blue) and $0.1$
(red and dashed) respectively.}
\end{figure}

In Fig. \ref{SLRealization}, we plot the charge density $\rho$ as a function of the chemical potential $\mu$ when $\langle{\cal
O_{-}}\rangle\neq0$ (left) and $\langle{\cal O_{+}}\rangle\neq0$ (right). For each chosen $\alpha$, we see that when $\mu$ is small, the system
is described by the AdS soliton solution itself, which can be interpreted as the insulator phase \cite{Nishioka-Ryu-Takayanagi}. When $\mu$
reaches $\mu_{- S}$ or $\mu_{+ S}$, there is a phase transition and the AdS soliton reaches the superconductor (or superfluid) phase for larger
$\mu$. Here we show that the phase transition can occur even at strictly zero temperature in the Gauss-Bonnet gravity, which is different from
that of the standard holographic superconductors with high curvature corrections  discussed in \cite{Gregory}.

\begin{figure}[H]
\includegraphics[scale=0.75]{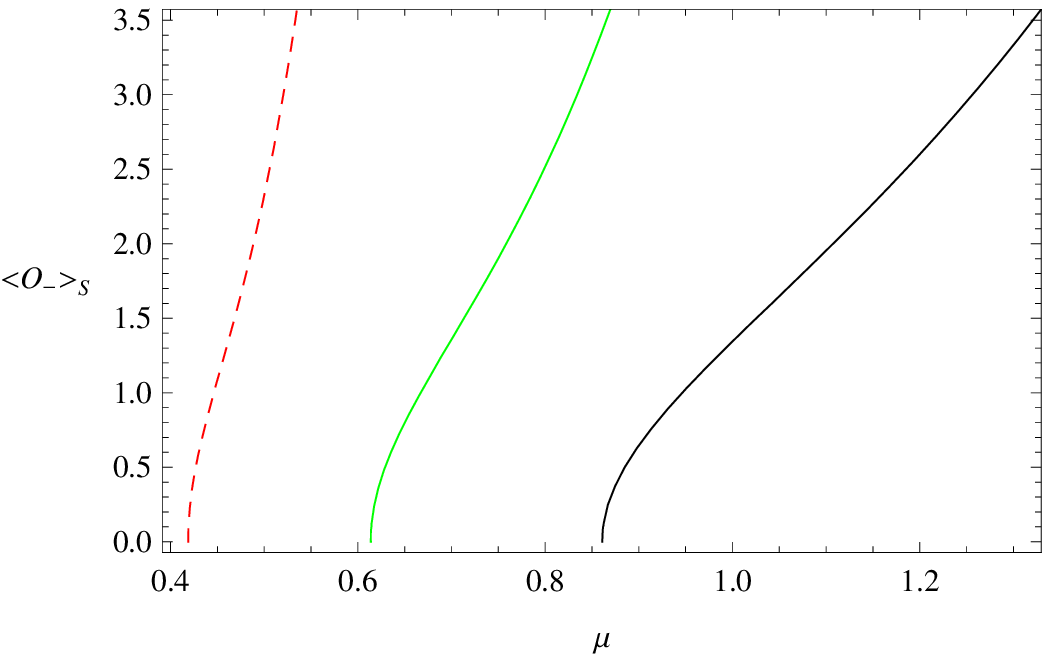}\hspace{0.2cm}%
\includegraphics[scale=0.75]{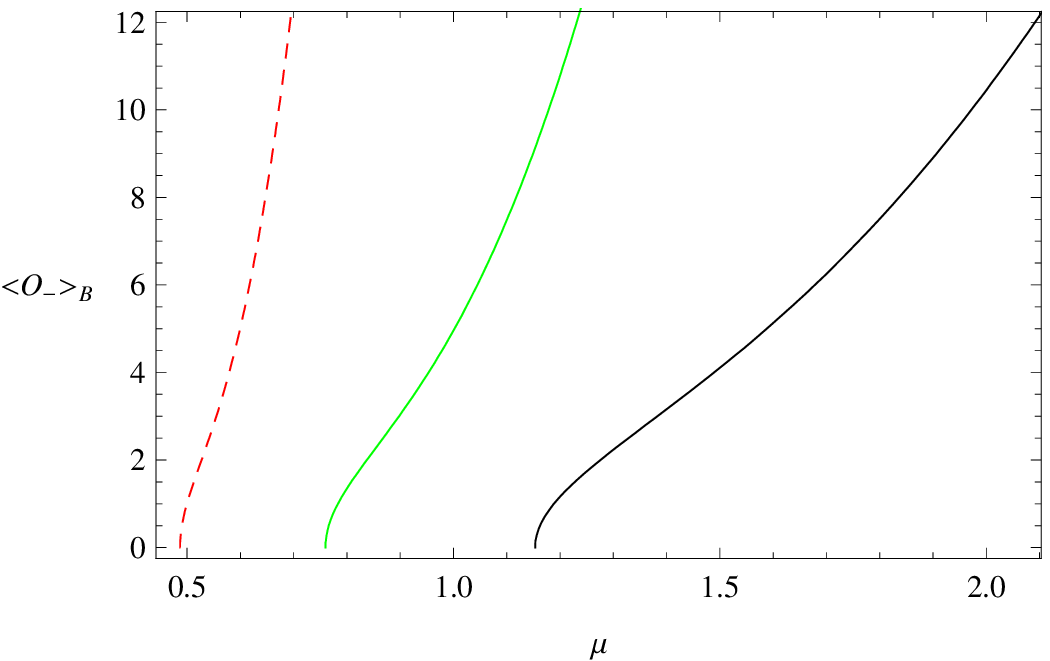}\\ \vspace{0.0cm}
\includegraphics[scale=0.75]{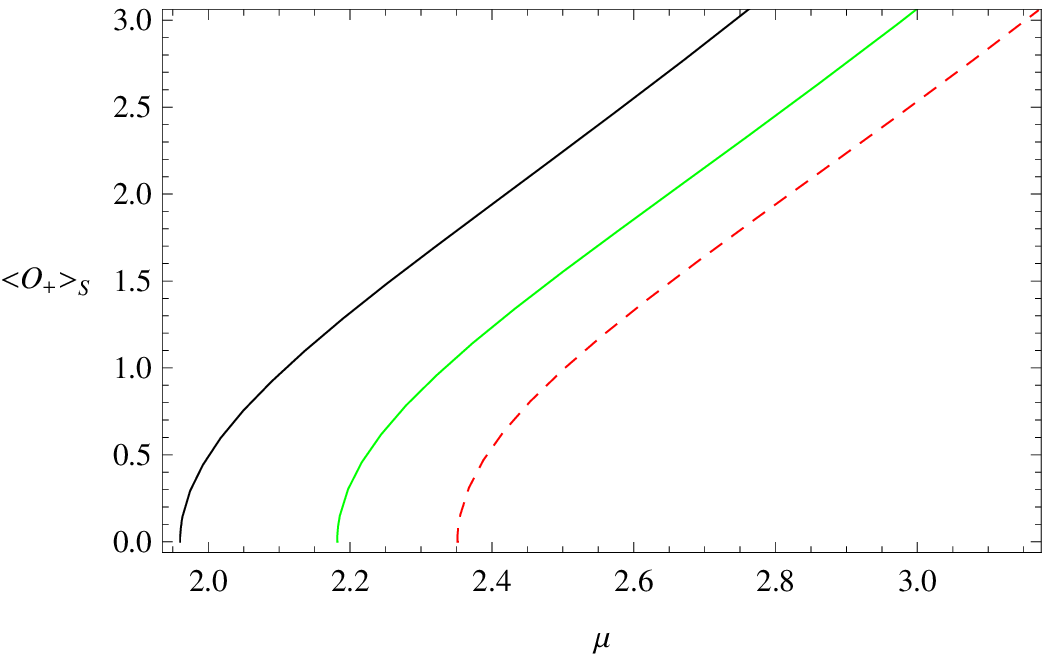}\hspace{0.2cm}%
\includegraphics[scale=0.75]{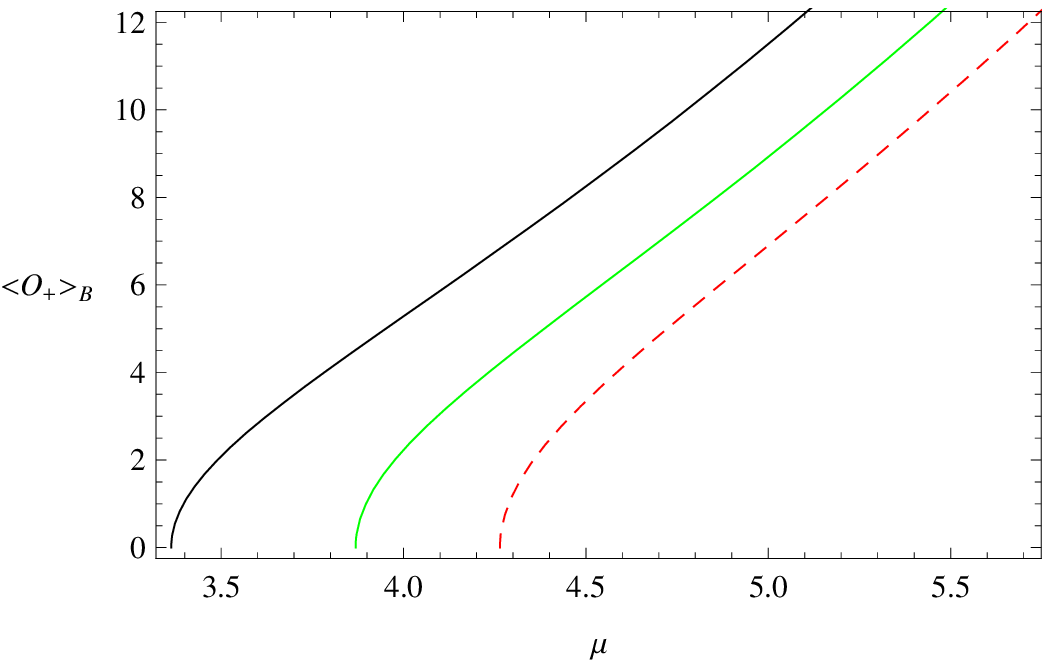}\\ \vspace{0.0cm}
\caption{\label{Cond-SolitonBHa01} (color online) The condensates of
the scalar operators $\langle{\cal O_{-}}\rangle$ and $\langle{\cal
O_{+}}\rangle$ with respect to the chemical potential $\mu$ in the
Gauss-Bonnet Soliton (left column) and Gauss-Bonnet Black hole
(right column) for $\alpha=0.1$. In the above two panels three lines
from left to right correspond to decreasing mass of the scalar
field, i.e., $m^{2}L_{\rm eff}^2=-13/4$ (red and dashed), $-7/2$
(green) and $-15/4$ (black) respectively, but they are arranged
contrarily in the following two panels.}
\end{figure}

\begin{figure}[H]
\includegraphics[scale=0.75]{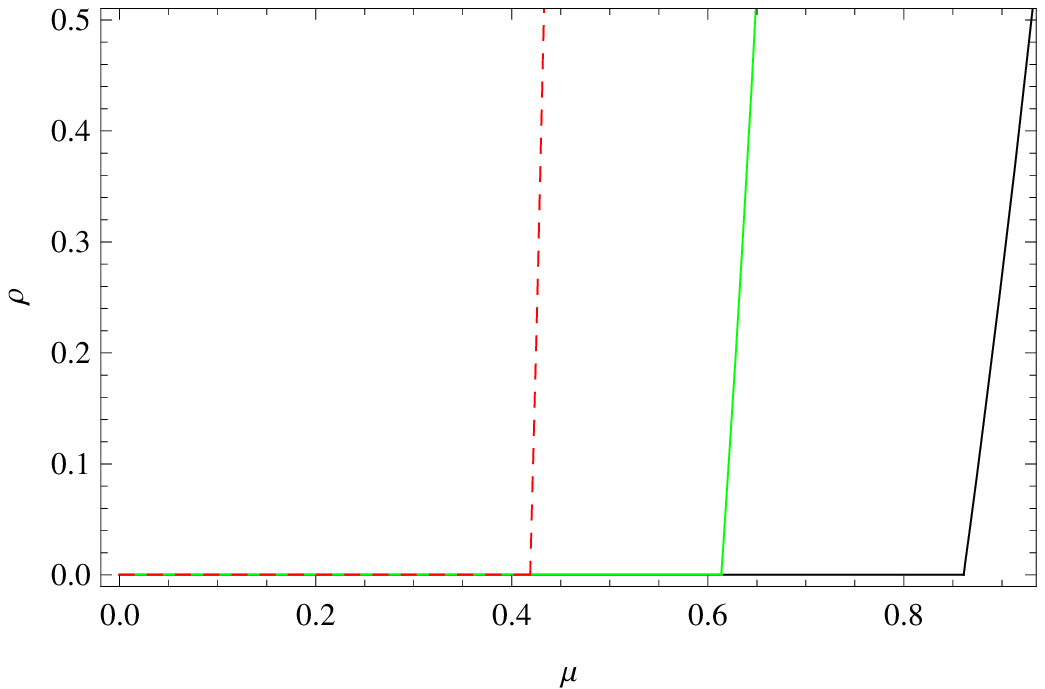}\hspace{0.2cm}%
\includegraphics[scale=0.75]{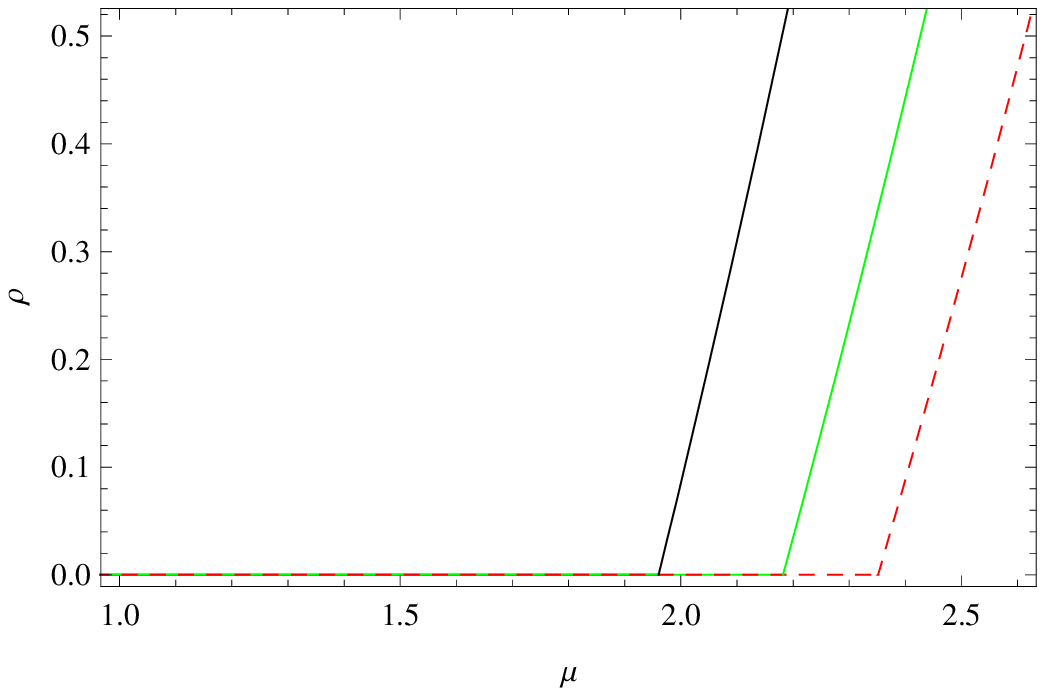}\\ \vspace{0.0cm}
\caption{\label{SLRealizationa01} (color online) The charge density
$\rho$ as a function of the chemical potential $\mu$ with different
mass of the scalar field when $\langle{\cal O_{-}}\rangle\neq0$
(left) and $\langle{\cal O_{+}}\rangle\neq0$ (right) for
$\alpha=0.1$. Their derivatives jump at the phase transition points.
The three lines in the left panel from left to right correspond to
decreasing mass of the scalar field, i.e., $m^{2}L_{\rm
eff}^2=-13/4$ (red and dashed), $-7/2$ (green) and $-15/4$ (black)
respectively, but they are arranged contrarily in the right panel.}
\end{figure}

In Fig. \ref{Cond-SolitonBHa01} we present the effect of the scalar field mass on the condensations of the scalar operators $\langle{\cal
O_{-}}\rangle_{S}$ and $\langle{\cal O_{+}}\rangle_{S}$.  Fixing the Gauss-Bonnet coupling, we observe consistent behaviors on the scalar mass
influence on the $\langle{\cal O_{-}}\rangle_{S}$ and $\langle{\cal O_{+}}\rangle_{S} $  comparable to the behavior of the AdS black holes. With
the increase of the scalar field mass, the critical chemical potential $\mu_{- S}$ (and $\mu_{- B}$) for the scalar operators $\langle{\cal
O_{-}}\rangle$ becomes smaller, however for the scalar operators $\langle{\cal O_{+}}\rangle$ where larger scalar filed mass leads higher $\mu_{+
S}$ (and $\mu_{+ B}$).  Fig. \ref{SLRealizationa01} shows the influence of the scalar mass on the behavior of the charge density as function of
$\mu$ for a fixed value of $\alpha$.

\subsection{Conductivity}

\begin{figure}[H]
\includegraphics[scale=0.75]{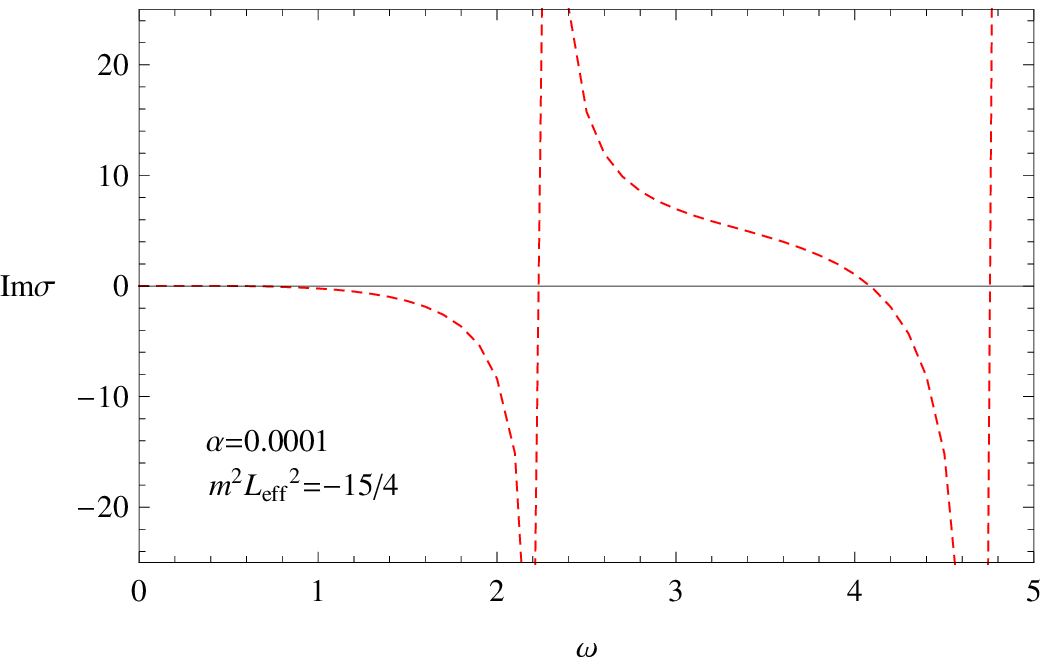}\hspace{0.2cm}%
\includegraphics[scale=0.75]{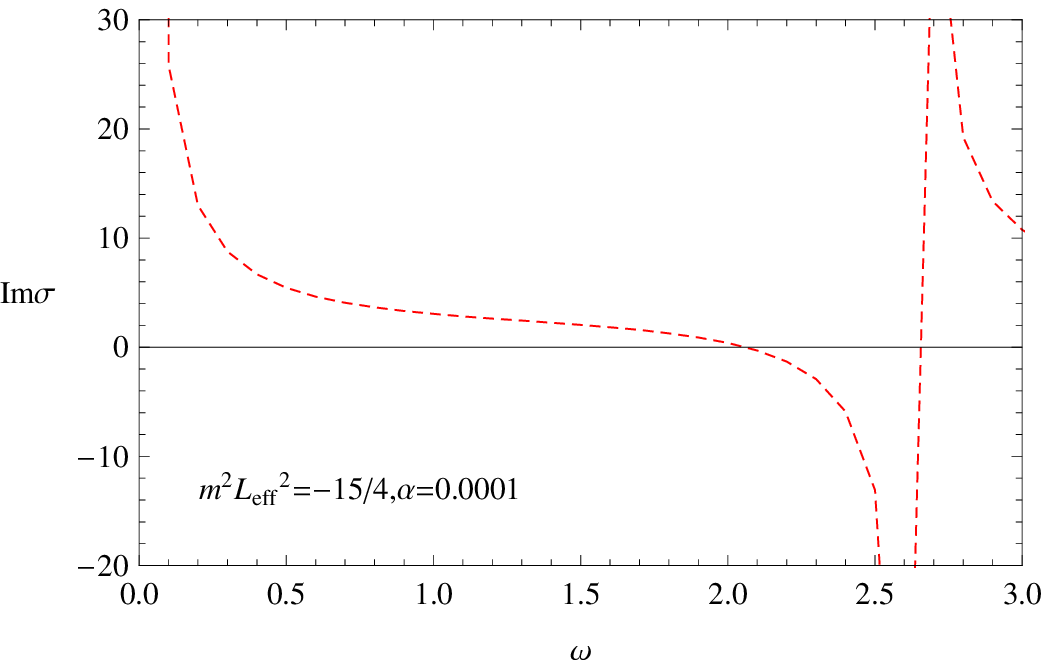}\\ \vspace{0.0cm}
\includegraphics[scale=0.75]{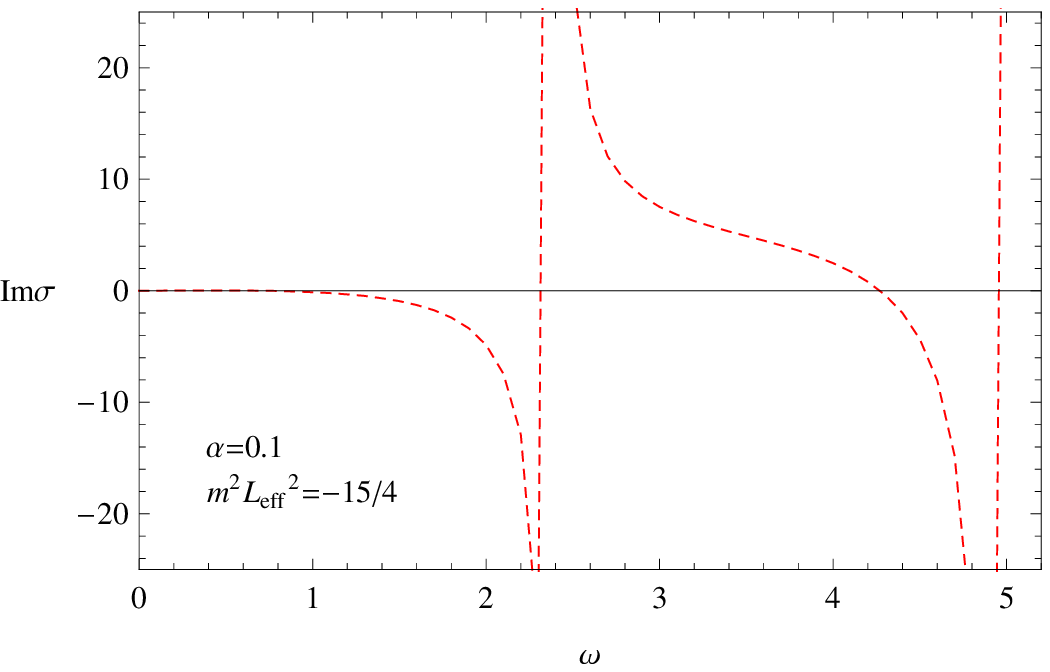}\hspace{0.2cm}%
\includegraphics[scale=0.75]{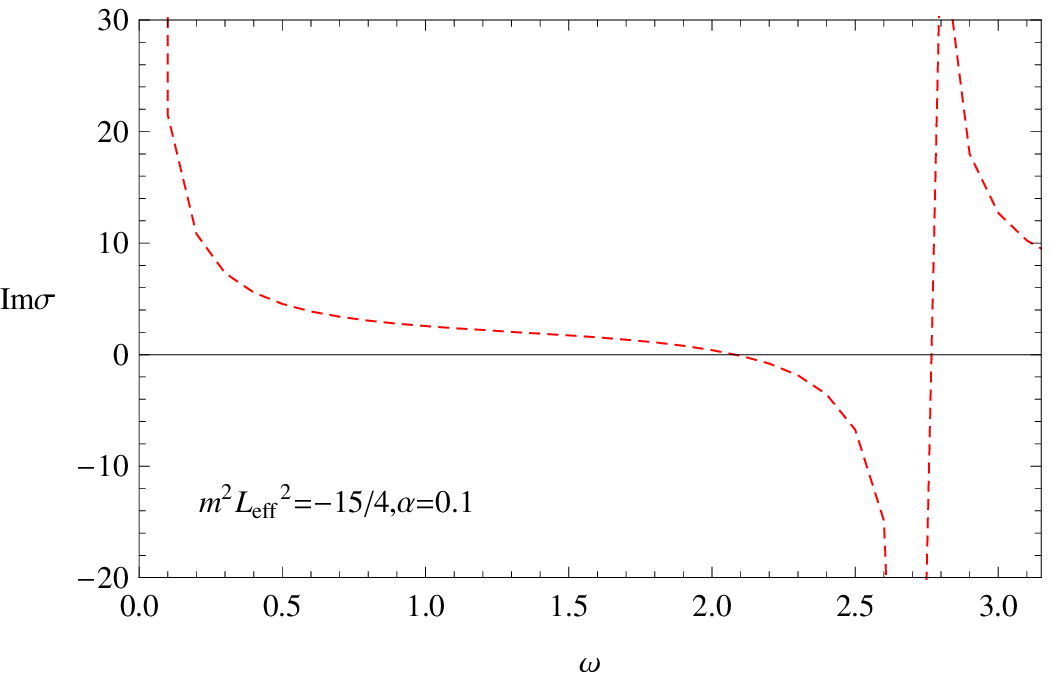}\\ \vspace{0.0cm}
\includegraphics[scale=0.75]{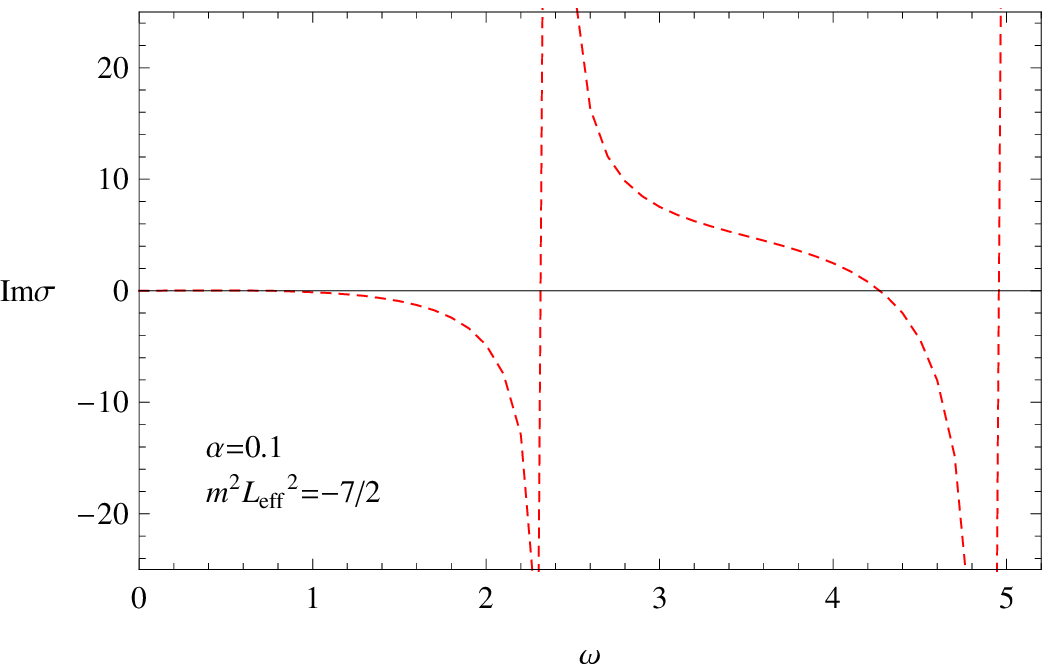}\hspace{0.2cm}%
\includegraphics[scale=0.75]{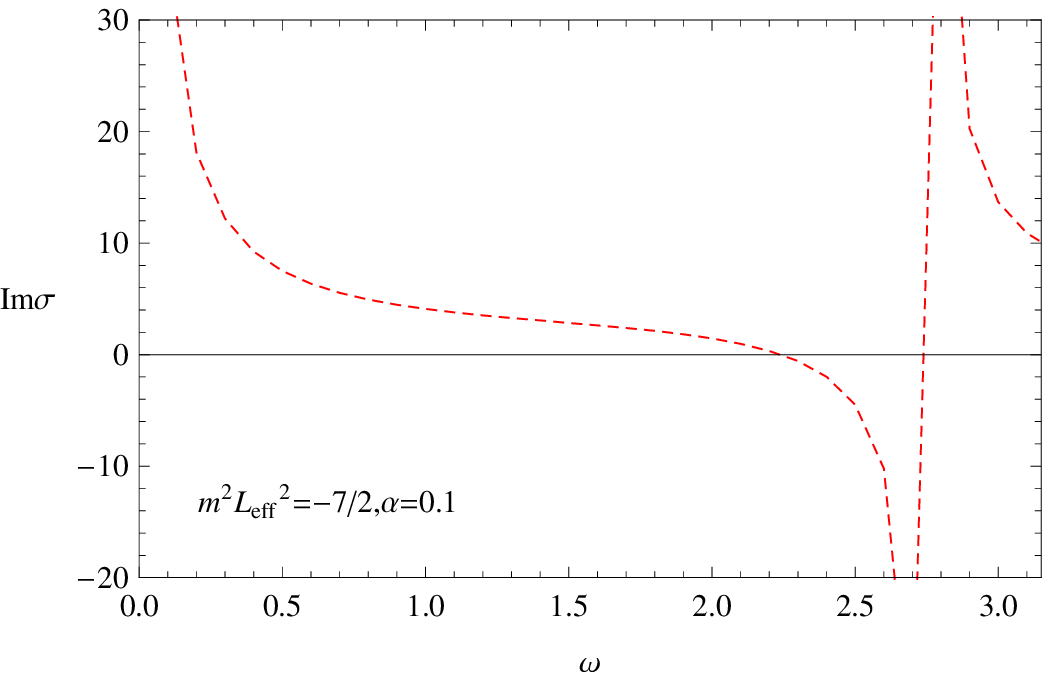}\\ \vspace{0.0cm}
\caption{\label{SLConductivity} (color online) The imaginary part of
the conductivity for the Gauss-Bonnet Soliton without a scalar
condensation $\langle{\cal O_{-,+}}\rangle=0$ (left) and with a
scalar condensation $\langle{\cal O_{-}}\rangle\neq0$ (right). We
choose the mass of the scalar field by $m^{2}L_{\rm eff}^2=-15/4$,
$-7/2$ and set $\alpha=0.0001$, $0.1$ for clarity.}
\end{figure}

We can calculate the conductivity $\sigma(\omega)$ by solving the equation for the electromagnetic perturbations $A_{x}$   in the AdS soliton
background for the Gauss-Bonnet gravity,
\begin{eqnarray}
A_{x}^{\prime\prime}+\left(\frac{f^\prime}{f}+\frac{d-4}{r}\right)A_{x}^\prime
+\left(\frac{\omega^2}{r^2f}-\frac{2\psi^2}{f}\right)A_{x}=0 \; ,
\label{Maxwell Equation}
\end{eqnarray}
which obeys the Neumann boundary condition as in Eq.
(\ref{SolitonBoundary}) at $r=r_{s}$.  The asymptotic behavior
near the boundary $r\rightarrow\infty$ is the same as
(\ref{Maxwell boundary}). The conductivity can be calculated by
using Eq. (\ref{GBConductivity}) and the results are shown in Fig.
\ref{SLConductivity}.

For clarity, we only plot the imaginary part of the conductivity for the Gauss-Bonnet-AdS Soliton without scalar condensation $\langle{\cal
O_{-,+}}\rangle=0$ (left) and with scalar condensation $\langle{\cal O_{-}}\rangle\neq0$ (right). For the Gauss-Bonnet-AdS soliton background
without scalar condensation, the AdS soliton can be identified with an insulator as argued in \cite{Nishioka-Ryu-Takayanagi}. There are periodic
poles at points where $A^{(0)}$ vanishes. With the increase of the curvature correction, the pole appears at bigger $\omega$. For the case of the
Gauss-Bonnet-AdS Soliton with scalar condensation as shown in the right column of Fig. \ref{SLConductivity}, we see that when $\omega$ is bigger,
the behavior looks similar to that in the left column. However the pole appears when $\omega=0$ which agrees with that observed for the
condensation in AdS black hole. For  larger Gauss-Bonnet coupling, we see that gap frequency in the imaginary part of the conductivity becomes
larger. This is in agreement with that in Gauss-Bonnet AdS black hole. For the fixed Gauss-Bonnet coupling constant, the influence of the scalar
mass on the imaginary part of the conductivity can be neglected.

\section{conclusions}

We investigated the behavior of a holographic superconductor in the presence of Gauss-Bonnet corrections to the gravity in the AdS bulk.
Considering that the Gauss-Bonnet term corresponds to the leading order string quantum corrections to gravity, this investigation may help to
understand the stringy effects to holographic superconductors. In the probe limit, we found that the mass of the scalar field and the
Gauss-Bonnet coupling influences the condensation formation and conductivity. In order to disclose the correct consistent influence due to the
Gauss-Bonnet coupling in various condensates, we found that it is more appropriate to choose the mass of the scalar field by selecting the value
of $m^2L^2_{eff}$,  since this choice contains directly the signature of Gauss-Bonnet factor in the scalar mass. The higher order curvature
corrections in general make the condensation harder to form while the increase of the dimensionality of the AdS space makes easier the scalar
operator to condense. To study the dynamics of the phase transition we used a semi-analytic method consisting in matching the solutions near the
horizon and the asymptotic AdS region at an intermediate matching point. We showed that this procedure breaks down in high dimensions $(d>5)$
unless the matching point is chosen in an appropriate range. We studied the ratio $\omega_g/T_c $ for various masses of the scalar field and
Gauss-Bonnet couplings and we found that the high curvature terms give large corrections to its universal value.

We also discussed a holographic superconductor dual to a
Gauss-Bonnet-AdS soliton.  Similar to that of the black hole
background, we observed that the condensation also appears in the
AdS soliton background. Although the Gauss-Bonnet term has no
effect on the Hawking-Page phase transition between AdS black hole
and AdS soliton, it does have effect on the scalar condensation
and conductivity in Gauss-Bonnet-AdS soliton configuration. We
found that the higher curvature correction makes it harder for the
scalar hair to form, which is similar to that seen in the black
hole background. We also observed that the  scalar mass has
similar effects to scalar condensation and conductivity as in the
AdS black holes  background.

There are still open problems to be investigated further. First of all, it is interesting to extend this work beyond the probe limit. In this
case we have also to address the problem of stability of the gravity backgrounds we are considering. Another interesting extension is to consider
Gauss-Bonnet black hole solutions with  spherical or hyperbolic horizons. Recently it was argued that the negative curvature topology can make
the superconductor gapless and give a geometrical mechanism of conductivity \cite{Koutsoumbas}. It is of interest to examine the spacial topology
influence on the condensation and work in this direction is in progress.

{\bf Acknowledgments}

We would like to acknowledge helpful discussions with E. Abdalla, R. G. Cai, R. K. Su and S. Y. Yin. This work was partially supported by the
National Natural Science Foundation of China.

\end{document}